\documentclass[a4paper,twocolumn,11pt,accepted=2023-01-27]{quantumarticle}
\pdfoutput=1
\usepackage[table]{xcolor}
\usepackage{float}
\usepackage[utf8]{inputenc}
\usepackage[english]{babel}
\usepackage[T1]{fontenc}
\usepackage{amsmath}
\usepackage{hyperref}
\usepackage[numbers,sort&compress]{natbib}

\usepackage{tikz}
\usepackage{lipsum}
\usepackage{bbm, amsthm, bm,textcomp, nicefrac,geometry,ragged2e}
\usepackage[bbgreekl]{mathbbol}
\usepackage{graphicx,epstopdf,color,verbatim,enumitem,ulem}
\usepackage{pbox,array}
\usepackage{mathrsfs}
\usepackage{amssymb}
\usepackage{mathtools}
\usepackage{stackrel}
\usepackage[thinlines]{easytable}
\usepackage{amsmath}
\makeatletter
\usepackage[caption=false]{subfig}
\usepackage[T1]{fontenc}
\usepackage[caption=false]{subfig}
\usepackage{babel}
\addto\captionsenglish{}

\usepackage{amsthm}
\usepackage{amsmath}
\usepackage{thmtools}

\usepackage[linesnumbered,ruled]{algorithm2e}
\usepackage{hyperref}

\newcommand\id{\mathbbm{1}}

\newcommand{\be}{\begin{equation}}
\newcommand{\ee}{\end{equation}}
\newcommand{\eea}{\end{eqnarray}}
\newcommand{\bea}{\begin{eqnarray}}

\newcommand{\ket}[1]{\left|#1\right\rangle}

\newcommand{\bra}[1]{\left\langle #1\right|}

\definecolor{brickred}{rgb}{0.8, 0.0, 0.0}

\begin{document}

\title{Optimized Quantum Networks}

\author{J. Miguel-Ramiro}
\affiliation{Institut f\"ur Theoretische Physik, Universit\"at Innsbruck, Technikerstra{\ss}e 21a, 6020 Innsbruck, Austria}
\orcid{0000-0001-9723-7298}
\thanks{jorge.miguel-ramiro@uibk.ac.at}

\author{A. Pirker}
\orcid{0000-0003-1260-1981}
\affiliation{Institut f\"ur Theoretische Physik, Universit\"at Innsbruck, Technikerstra{\ss}e 21a, 6020 Innsbruck, Austria}
\author{W.~D\"ur}
\affiliation{Institut f\"ur Theoretische Physik, Universit\"at Innsbruck, Technikerstra{\ss}e 21a, 6020 Innsbruck, Austria}
\orcid{0000-0002-0234-7425}

\maketitle

\begin{abstract}
The topology of classical networks is determined by physical links between nodes, and after a network request the links are used to establish the desired connections. Quantum networks offer the possibility to generate different kinds of entanglement prior to network requests, which can substitute links and allow one to fulfill multiple network requests with the same resource state. We utilize this to design entanglement-based quantum networks tailored to their desired functionality, independent of the underlying physical structure. The kind of entanglement to be stored is chosen to fulfill all desired network requests (i.e. parallel bipartite or multipartite communications between specific nodes chosen from some finite set), but in such a way that the storage requirement is minimized. This can be accomplished by using multipartite entangled states shared between network nodes that can be transformed by local operations to different target states. We introduce a clustering algorithm to identify connected clusters in the network for a given desired functionality, i.e. the required network topology of the entanglement-based network, and a merging algorithm that constructs multipartite entangled resource states with reduced memory requirement to fulfill all desired network requests. This leads to a significant reduction in required time and resources, and provides a powerful tool to design quantum networks that is unique to entanglement-based networks.
\end{abstract}

\section{Introduction}
Networks play an important role in many branches of science, ranging from biology over social sciences to communication. Their properties and emerging effects are determined by their topology, which in turn is given by the physical setting. This is e.g. the case in classical communication networks such as the internet, where nodes are connected in a specific way by channels that correspond to available fibres or free-space communication links. Also a quantum network \cite{Kimble2008,Kozlowski2019,Azuma2021} has an underlying physical structure and topology, which is used to enable quantum communication between nodes. However, such a quantum network can also be used to establish multipartite entangled states, allowing for the design of entanglement-based networks \cite{Pirker2018,Pirker2019,Meignant2019,Gyongyosi2019,delocalizedinfo} where the functionality and topology is provided by pre-established entangled resource states that are used as a resource to fulfill network requests, including e.g. teleportation of quantum information \cite{Bennett_telep}, or perform other tasks such as remote sensing \cite{Eldredge2018,Sekatski2020} or distributed quantum computation \cite{CiracDistributed,Cacciapuoti2020}. 

Here we show that one can use this principle to design quantum networks with optimized topology, independent of the underlying physical structure. This allows one to overcome restrictions and bottlenecks inherent in physical networks, and adjusting any given network to the desired functionality by properly choosing the resource states to be stored. Such an approach is unique to quantum networks, as resources for quantum communication can be established prior to the announcement of a specific request. We are interested in the optimal entangled states to be stored, where we consider the memory requirement, i.e. the total number of qubits that need to be stored in the network, as a natural figure of merit.   

We take the desired functionality of the network as a starting point, i.e. we specify which requests a network should be able to fulfill. A request could be e.g. the generation of one or several (parallel) Bell pairs shared between some selected nodes in the network that can serve as quantum communication channels, or the generation of some multipartite entangled target state.
A network state should be able to handle all possible requests, i.e. one should be able to transform the network entangled state shared by the nodes into each possible request configuration with local operations. Throughout this work we restrict requests to multiple parallel bipartite connections. We utilize three different strategies to construct optimal network resource states:
\begin{itemize} 
\item We provide a clustering algorithm that allows a hierarchical structure within the network. The algorithm identifies strongly connected clusters that are only weakly connected among each other. Clusters are identified with sub-networks, and treated independently. Connections between clusters are considered at the next layer, where each cluster is represented by a single node or router, with resource states shared among routers.
\item We introduce a merging algorithm that allows one to identify a multipartite resource state with the same functionality as a collection of different resource states, but with smaller memory requirements.
\item We propose generalized constructions for efficient resource states that allow for arbitrary simultaneous pairwise connections.
\end{itemize}
Each of these approaches itself leads to a significant reduction in memory requirements as compared to simply storing all states that might be requested. A combination of the different approaches, applied in a hierarchical way at all layers, leads to enhanced  entanglement based networks with reduced memory requirements that outperform techniques based on bipartite entanglement only. Notice that the clusters do not respect the initial geometry of the network, but are solely determined by the required functionality. The underlying geometry, i.e. the physical connections between nodes, influence the difficulty and cost of establishing different kinds of entangled states shared among the nodes. This corresponds to the dynamical phase \cite{Pirker2018,Pirker2019}, and in a standard approach to quantum networks \cite{Matsuzaki2010,Meter2011,Epping_2016,Wehner2018,Pirandola2019} where entanglement is built-up upon request, this is usually the main challenge that is addressed. 

Here, however, we take a different approach and argue that resource states can be established prior to requests, i.e. in times when the network would be idle. Then the key challenge is to identify suitable entangled resource states that allow one to fulfill (or assist) any request, without transmitting any further quantum information. As these states need to be stored --static phase--, and finally adapted by means of local operations at the different nodes to fulfill the required request --adaptive phase-- \cite{Pirker2018,Pirker2019}, the required memory for storage is the central quantity of interest. Such an approach has the additional advantage that the time-costly generation of long-distance entanglement does not start when a request arrives, but has been done beforehand, thereby speeding up the fulfillment of a request significantly. 

\section{Preliminaries}
\subsection{Network states and transformations} \label{sec:basicops}
In this work we consider entangled-based networks where a resource or network state is initially shared between all the network nodes. These resource states involve multipartite entanglement that can be in general described by graph states. We do not consider the preparation process of the resource state, but only its storage. On the other hand, here we mostly restrict to communication requests that only involve bipartite links between parties, i.e. target states are Bell states. 

Bell states are two qubit maximally entangled states that can be represented as
\begin{align}
\ket{B_{ij}} = (\id \otimes \sigma_{x}^j \sigma_{z}^i) \ket{B_{00}}, \label{eq:bellbasis}
\end{align}
with $i, j \in \{0, 1\}$, and where $\sigma_{x}$ and $\sigma_{z}$ are the $X$ and $Z$ Pauli operators. These states play an important role with different applications in e.g. teleportation \cite{Bennett_telep}, quantum key distribution \cite{Ekert91} or superdense coding \cite{Bennett_supdencoding}. 

The most natural extension of these maximally entangled states to multipartite (more than 2 parties) systems are GHZ states, defined as
\begin{align}
	\ket{\mathrm{GHZ}_n} = \frac{1}{\sqrt{2}}\left( \ket{0}^{\otimes n} + \ket{1}^{\otimes n} \right),
\end{align}
which represent one particular type of graph state (up to local basis change).

Graph states $|\psi_G\rangle$ \cite{Hein2004,Hein2006} are a certain subclass of mutipartite quantum states where the states are associated and can be represented as graphs $G=(V,E)$. Qubits are associated with vertices of the graph, and the corresponding graph state $|\psi_G\rangle$ is defined as the unique $+1$ eigenstate of the stabilizers
\begin{align}
K_{a} = \sigma_{x}^{(a)} \prod_{(a,b) \in E} \sigma_{z}^{(b)}, \label{def:graph}
\end{align}
for all $a \in V$. Equivalently, graph states can be described by performing a controlled-Z gate $CZ=diag(1,1,1,-1)$ between any two qubits that are connected by an edge, i.e.
\begin{align}
|\psi_G \rangle = \prod_{(a,b) \in E} CZ^{(a,b)} |+\rangle^{\otimes N},
\end{align}
where $|+\rangle =(|0\rangle + |1\rangle)/\sqrt{2}$ is the +1 eigenstate of $\sigma_x$.

Notice that graph states include genuine mutipartite entangled states with different entanglement features, but also multiple smaller, disconnected graph states shared between different subsets of the nodes, i.e. $|\psi_G\rangle=\bigotimes_j |\psi_{G_j}\rangle$. One such instance of particular importance are multiple Bell states shared between different nodes. This corresponds to a graph with multiple, separated edges, i.e. each node appears at most once in an edge, where each Bell pair is represented by two nodes connected by an edge.

Graph states exhibit interesting manipulation and transformation properties under the effect of certain quantum operations \cite{Hein2004,Hein2006}. Among these properties, we can highlight:

\textit{Vertex deletion.---} A vertex can be removed by applying a Pauli Z measurement on that qubit, up to local correction operations. All the edges associated with this vertex are deleted as well.

\textit{Local complementation.---} Given some vertex of the graph, its local complementation \cite{Hein2006} inverts the subgraph corresponding to its neighborhood, leading to an equivalent graph state. A measurement in the Y basis produces a local complementation and deletion of the measured vertex, whereas a measurement  of a qubit in the X basis corresponds to a local complementation of a neighbor of the measured qubit, followed by locally complementing the resulting graph w.r.t. the measured qubit \cite{Hein2006}. In both cases the qubit and the edges associated are deleted after the measurement.

\textit{Merging operations.---} Two qubits (vertices) that belong to the same or to different graph states can be merged either using a Bell measurement, such that both vertices are deleted and their neighborhoods connected or using a merging measurement \cite{MiguelRamiro2021}, which merge both vertices into a single one. This last operations simply consists of a measurement of two qubits $a_{1}, a_{2}$ described by $\{P_{0},P_{1}\}$, with 
\begin{align}
&P_0 = \ket{0}_{a}\bra{00}_{a_{1}, a_{2}} + \ket{1}_{a}\bra{11}_{a_{1}, a_{2}}, \notag \\
&P_1=\ket{0}_{a}\bra{01}_{a_{1}, a_{2}} + \ket{1}_{a}\bra{10}_{a_{1}, a_{2}}, \label{eq:merging}
\end{align}
resulting in a merged qubit $a$.
All these operations, applied in a local way (local w.r.t. each node), should allow the network to transform the resource state into each of the desired target states. 

We assume that target states (or requests) are graph states $|\psi_{G}\rangle$. We will however mainly restrict to target states corresponding to multiple Bell states, i.e. multiple parallel bipartite connections between different nodes, where each node takes part only in at most a single connection per request. We will also consider the transformation of a given resource state to one out of many possible target states. 

A simple example of a multipartite resource state that can be transformed by local operations to different target states is given by the GHZ state. Measuring all but two qubits in the $\sigma_x$ basis leaves the two remaining qubits in a Bell state. So a $N$-qubit GHZ state can be transformed to a Bell state shared between any two nodes \cite{Hein2004}. 

Another relevant example consists in establishing a direct connection (Bell state) between any two nodes $a,b$ that belongs to a connected graph state. A direct approach, denoted as repeater-path protocol, consists in isolating a path between them by measuring in the Z basis the neighborhood of the path, followed by X measurements of the path qubits. Alternatively, one can also first measure the path qubits in the X basis followed by Z measurements of the remaining neighborhood, achieving the desired connection with --in general-- less measurements (see \cite{Hahn2019}). 
A central tool in this work involves deciding whether a given graph state can be transformed into another one --or a collection of Bell pairs-- by local operations. This has been proven to be in general a NP-Complete problem \cite{Dahlberg2018,Dahlberg2020}, which make the optimization problem we treat here a complicated task (see below).

In addition to this, we make use of different tools or techniques from classical graph theory \cite{GraphTheory}. In particular, we are interested in connectivity properties of graphs. A crucial tool in this respect is the graph laplacian \cite{Das2004}, defined as the difference between the degree and the adjacency matrices (see below for details), whose spectrum provides relevant information about the graph connectivity and its homogeneity. Based on this, one can group network elements with similar connectivity among them by using well known techniques of spectral clustering \cite{vonLuxburg2007}, which takes advantage of the information provided by the eigenspectrum of the laplacian in order to group the different nodes. We adapt and extend these tools for our purposes in a quantum network scenario.

\subsection{Relation to previous work}
Quantum networks constitute an important and active field of research. In particular, studying their design and architecture have received significant attention lately. Many of these approaches rely on the use of quantum repeaters \cite{Br98,DurRepeater}. Denoted as quantum repeater networks, they make use of the repeater to recursively construct network states dealing with noise and losses that allow for communication over long distances \cite{Meter2013a,Meter2013b,Muralidharan2016}. Bounds and optimal capacities for quantum communication and entanglement distribution in repeater networks have been analyzed \cite{Pirandola2017,Pirandola2019}. Similarly, network states can be built by using multipartite approaches, where the task derives in optimally distributing these multipartite entangled states --and graph states in particular--, also dealing with noise and imperfections \cite{Matsuzaki2010,Cuquet2012,Epping_2016,Epping_2016_2,Pirker2018,Pirker2019,Khatri2019}, in such a way that they can overcome the bipartite approach by exploiting multipartite entanglement properties \cite{Epping2017,Hahn2019}.

Fulfilling quantum communication requests, i.e. quantum routing, such as establishing certain bipartite or multipartite states among certain nodes of the network, is another important research direction. Closely related to entanglement distribution, two directions can be followed. On the one hand, a bottom-up approach \cite{Meter2013b,Schoute2016,Gyongyo2017,Gyong2018,Pant2019,Chakr2019} where, depending on the task, the network makes use of the local resources to fulfill the request. Alternatively, one can follow a top-down approach \cite{Pirker2018,Pirker2019,Chakr2019}, where the network is first provided by a resource state, which is stored and used subsequently to fulfill the requests by applying certain local operations on the different network nodes. 

Notice that the scenario we treat relies on pre-shared entanglement. However, we do not explicitly treat the entanglement generation, i.e. the dynamic phase, but consider only the static phase for storage, and the adaptive phase to generate the desired target configurations. In this we differ from many previous works that are mainly concerned about establishing entangled states in the network. Our approach is also compatible with e.g. bipartite repeater networks, since bipartite entanglement can be established and multipartite entanglement can be subsequently created by suitable merging processes in order to optimize network storage. We consider a top-down approach where the entanglement is provided beforehand, establishing a network state, and all the network requests are known beforehand, such that the network state has to be able to guarantee any of them.

\section{Setting}
We introduce in this section the general setting that we consider throughout this work. This setting is conceived in such a way it can represent an hypothetical but realistic network scenario. The network users can require different connections or desired target states they want to share among each other, and our task is to provide a single resource network state that can guarantee --via local operations only-- any of the requests (see also Appendix \ref{Sec:Appendixswitch}).

{\bf Network request}: The $j^{\rm th}$ target configuration or network request is described by a graph $G_j$, where the corresponding graph state $|\psi_{G_j}\rangle$ should be established. Here $j \in \{1,\dots, m\}$ with $m \in \mathbb{N}$.

{\bf Network functionality:} The desired functionality of the network is described by the set of all possible requests, where we consider that each configuration occurs with a given probability. Hence the functionality of the network is described by the set $\{(p_j,|\psi_{G_j}\rangle)\}$. 

\textit{Deterministic case.---} We consider two different scenarios. The first is a deterministic, single use scenario where any of all possible requests has to be fulfilled deterministically. That is, the resource state needs to be
such that it can be transformed to any one of the graph states $\{|\psi_{G_j}\rangle\}$, and we are interested in a single use of the network only. Notice that single use actually refers to many parallel links that are established simultaneously, but each node is part of only a single state. Probabilities are irrelevant in this case, and we consider them to be equal without loss of generality. 

For simplicity, in the remainder of this work we consider that each of the the target requests involves only multiple, separated bipartite connections between different nodes. This implies that each node is part of at most one connection, and all target configurations correspond to tensor products of Bell states. 

The task is to optimize an entangled-based network state that can be transformed with local operations into each of the possible target configurations. We are interested in minimizing the storage of the network or resource state, i.e. minimize the number of qubits to be stored at the different sites. 
This network state may consist of several multipartite entangled graph states shared between different nodes, where each node can hold multiple qubits. That is, we consider a resource state of enlarged size that allows one to fulfill all possible requests with local operations, and guarantee the full desired functionality of the network. One possible solution is to simple take the union of all target states that appear in any of the requests. A trivial size reduction is achieved by adding only target states that are not already present in any of the previous requests. 

However, there are also certain multipartite resource states that allow for a further size reduction as we discuss in more detail below. One example was pointed out above: A GHZ state of $N$ qubits suffices to fulfill all requests that only involve a single Bell pair between two nodes as target state. Hence a single GHZ state guarantee a network functionality to establish a single Bell pair between any pair of parties, where the network request specifies which parties want to communicate. There are $N(N-1)/2$ possible requests of this kind. The trivial construction of adding all possible target states would lead to a resource state of size $N(N-1)$, a tensor product of all possible Bell pairs. So we encounter a reduction from size $N(N-1)$ to $N$ w.r.t. qubits to be stored. 

In case the request configurations involve several simultaneous connections, different solutions has to be designed. For instance, we show that for the full pairwise connectivity case, i.e. any --combination of-- simultaneous links has to be guaranteed,  a configuration of $N-1$ Bell states, each shared between a central station and the other parties (i.e. $2(N-1)$ qubits to be stored) suffices --and is optimal at least for small networks-- if one allows for local operations. More general situations require more general approaches, for which we introduce network clustering and merging protocols. 

Although the requests have to be guaranteed with local operations, an important feature differentiates the solutions we propose in this work, i.e. 

{\bf Local operations}: We refer to local operations as any operation that is performed within a node. If the node stores more than one  qubit, these operations may include multi-qubit operations, such as e.g. merging measurements (Sec. \ref{sec:basicops}). 

{\bf Single-qubit operations}:  This represents a specific case within the local operations, where the request can be fulfilled by only involving operations on each qubit that a node possesses, independently of the number of qubits the node stores. These operations are in general easier to realize and introduce less amounts of noise.

\begin{figure}
\includegraphics[scale=0.25]{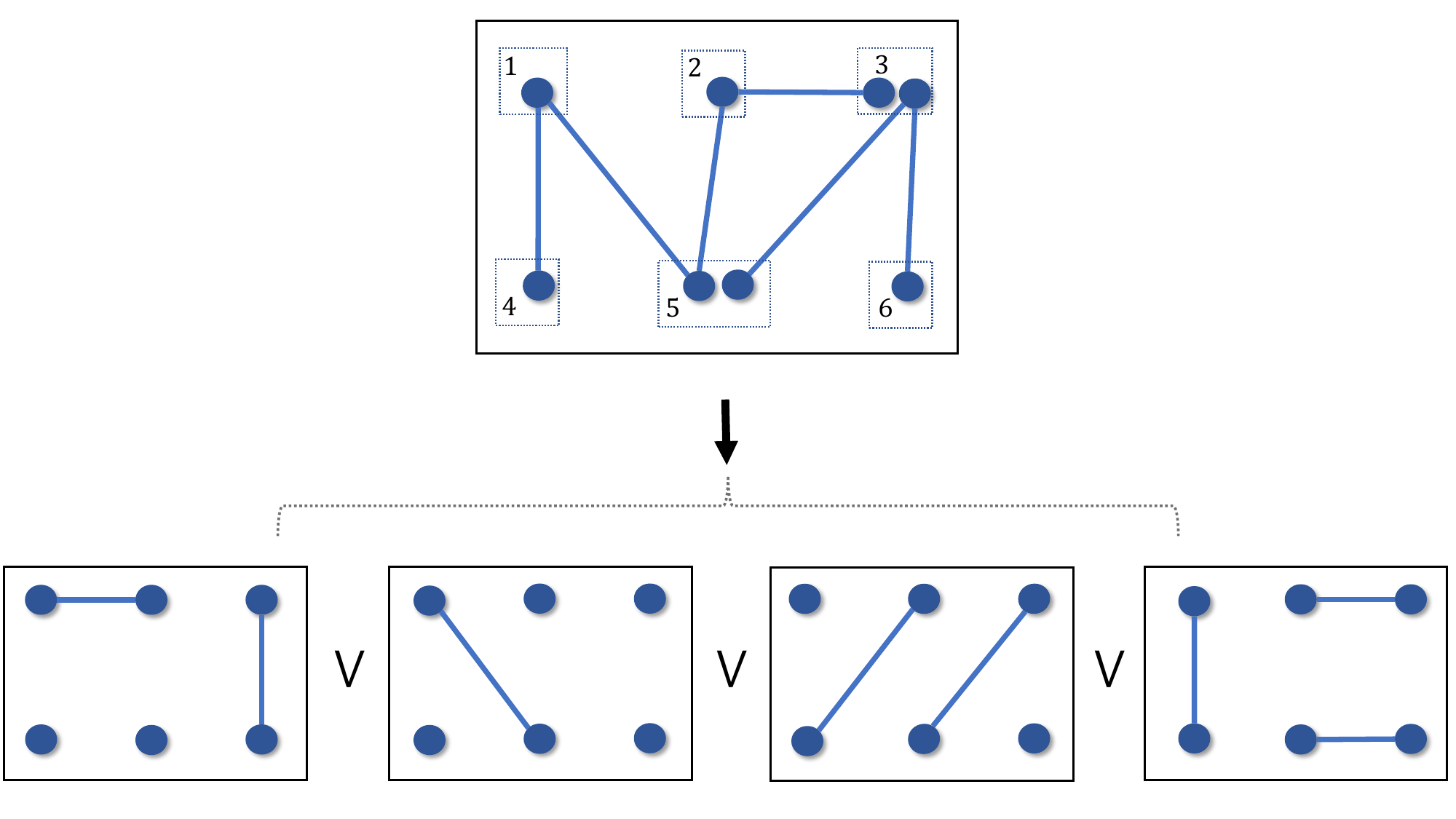}
\caption[h!]{\label{fig:Settingpicture} Setting example with a $6-$qubits network. Different requests are required and an entangled-based resource state is provided such that it can guarantee any request with the minimum possible number of stored qubits (see text).}
\end{figure}

Fig. \ref{fig:Settingpicture} shows a simple example with $6$ nodes and $4$ different requests. The network state, consisting of a 1D cluster state $\ket{C}_{4,1,5,2,3}$ and a GHZ state $\ket{GHZ}_{3,5,6}$, can guarantee any of the $4$ requests with local complementations and $Z$ measurements (vertex deletion). For instance, in order to achieve the last request (right), the following operations are needed: $Z_{5_{\text{l}}} Z_{3_{\text{r}}} LC_{3_{\text{r}}}$, where $r$ and $l$ indicates right and left qubit within each node respectively.

\textit{Probabilistic case.---} The second scenario takes the different probabilities for requests into account, and is concerned with multiple usages of the network, i.e. the resource state has to guarantee $k$ request calls, up to certain failure probability. A direct solution consists in providing $k$ copies of the resource state derived for the deterministic scenario. This can be however too demanding, and allowing for certain failure probability $p_{\rm succ}=1-\epsilon$ better solutions can exist that lead to significant storage savings.

In this case, one considers the average number of qubits that need to be stored in order to fulfill all possible requests, again restricted to local operations and single-qubit operations. One may either consider only a limited number of requests, or the (asymptotic) limit of many requests calling rounds.

\subsection{Remarks on resources and optimality} \label{Sec:Optimality}

It is important to discuss and specify the concept of optimality in the context we deal with. On the one hand, recent works (see e.g. \cite{Dahlberg2018,Dahlberg2020}) have stressed the difficulty of finding optimal solutions to the problems we deal with. Therefore, the task of finding optimal solution should be in general replaced by the task of finding good solutions. Moreover, one should define precisely the parameters or properties to be optimized. We briefly discuss these two points in the following.

Given some entangled resource state, one has to decide whether it can be transformed into any of the target configurations by using local operations. This involves single-qubit Clifford operations (LC), single-qubit Pauli measurements (LPM) and classical communication (CC) - and in some cases also multi-qubit operations that are local, i.e. only involve qubits at a given node. This closely relates with the problem of deciding whether a certain graph state can be transformed into another one. This has been analyzed in detail, and as shown in \cite{Dahlberg2018},  it implies in general a NP-Complete problem. Analogously, it has been also proven \cite{Dahlberg2020} that the problem of deciding whether a certain graph state can be transformed into a collection of Bell pairs by LC+LPM+CC operations is also NP-Complete. It is therefore not realistic to aim at finding optimal solutions for these tasks. In addition, results of these works indicate that building the resource states from the target configurations --and not the other way around-- could be a smarter approach.

On the other hand, one has to define the variable(s) or parameter(s) with respect to which to optimize. In this work, the quantity to be optimized is the quantum memory, i.e. the total number of qubits that each network node needs to store to guarantee the desired functionality. This choice is motivated by the fact that the resource network state has to be build and established for each request (or equivalently multiple copies have to be provided beforehand), and then be stored until it is used. System storage is a relevant (and in the near future limited) quantity whose reduction implies protocol advantages. Besides we remark that, although in a less explicit way, our solutions also provide improvements in terms of the amount and stability of the entanglement required by the network resource states.

\subsection{Tools for network state optimization} \label{Sec:Setting details} 

We detail here the mathematical and graphical tools that we use later. We assume that all the target configurations are known. 

\textit{Adjacency matrix.} In analogy to graph theory \cite{GraphTheory}, the adjacency matrix $A$ is a $n \times n$ symmetric matrix where each element
\begin{equation}
A_{i,j}={0,1} 
\label{eq:adjmatrix}\end{equation}
indicates whether nodes $i$ and $j$ are connected by an edge. In our case we denote $A_{i,j}={1}$ if nodes $i$ and $j$ are connected in any target configuration, independently of which one.

\textit{Simultaneous matrix.--} We introduce the simultaneous matrix $S$ as a weighted adjacency matrix where each element $S_{i,j}$ indicates how simultaneous, together with other links in the same request, is the connection $S_{i,j}$. In order to build this matrix, one takes each target configuration ${q} \in \{1,...,m \}$, where $m$ is the total number of configurations, and assign to every nonzero edge $s^{q}_{i,j}>0$  a new value or weight corresponding to $s^{q}_{i,j}=\sum_{l,r}A_{l,r}^{q}$, i.e. the number of edges within the $q$ configuration. Finally, all the target configurations are considered to assign a value
\begin{equation}
S_{i,j}=\max_{{q} \in \{1,...,m \}} s^{q}_{i,j}
\label{eq:simultaneousmatrix}\end{equation}
to each element of the simultaneous matrix. 

\textit{Cumulative matrix.--} We define the cumulative matrix $C$ as a weighted adjacency matrix where each element $C_{i,j}$ indicates the total number of requests in which a link appears, i.e.
\begin{equation}
C_{i,j}=\sum_{q=1}^{m} A_{i,j}^{q}.
\label{eq:cumulativematrix}\end{equation}
This matrix gives information about how often each connection occurs.


\textit{Graph laplacian.--} We define a graph laplacian operator $L$ as a function of the cumulative matrix $C$ introduced above, such that:
\begin{equation}
L=D-C, \label{eq:laplacian}
\end{equation}
where $D$ is the weighted degree matrix associated to the cumulative matrix $C$, i.e. a diagonal matrix where each element $D_{i,i}$ is given by the sum over the weights of all incident of vertices of node $i$. The spectrum of this laplacian operator $L$ contains information of how the connectivity comprising all the network requests is distributed, as a function of the number of occurrences of each link. Note that our definition of the laplacian differs from the one of graph theory, where it is typically defined as a function of the adjacency matrix \cite{Das2004,vonLuxburg2007}.

\section{Memory and entanglement optimization protocols} \label{sec:fullconecprotocols}

In order to design the resource network state one can follow different directions. We introduce here three independent strategies. First, we propose the application of clustering techniques from graph theory \cite{GraphTheory,Das2004,vonLuxburg2007} that we modify and adapt to our setting. This technique successfully groups nodes with similar connectivity, and allows us to identify subnetworks whose functionality can then be guaranteed by specific network states of reduced size. Subnetworks can then be replaced by a single router node, which are interlinked via networks resource states. This grouping can be hierarchically employed, thereby identifying different layers of a network. Notice that in contrast to hierarchical classical networks, here the subnetworks and layers are chosen solely by the desired functionality, and without the need to take the restrictions due to the underlying topology of physical links into account.

Second, we consider multipartite resource constructions that guarantee simultaneous pairwise connectivity of multiple Bell pairs shared between different nodes. Since all permutations of such configurations can be achieved, the corresponding network resource states are fully flexible, and offer a significantly reduced size as compared to including all possible target states in the network state. We present different constructions to achieve this aim. However, in many situations they require more resources than a tailored state that can only achieve a limited number of target configurations. 

In our third approach, we introduce a merging algorithm that takes into account all requests appearing in the desired network functionality, and construct from there a multipartite state of reduced size. In this approach we start by considering the union of all states appearing in the different requests, and systematically reduce the size of this network resource state by merging multiple states in such a way that they resulting state can still be transformed to all desired target configurations. While this does not necessarily lead to an optimal state, we demonstrate that in many situation there is a significant reduction of the required memory - both compared to the union of all target states, and w.r.t. full connectivity constructions of the second approach. Moreover, only single-qubit operations are required to transform the resource state into any of the requests.

Notice that the three approaches are independent of each other, and the clustering technique can be combined with the second and third approach.

\subsection{Network state clustering} \label{sec:clusteriing}
\begin{figure}
\includegraphics[scale=0.25]{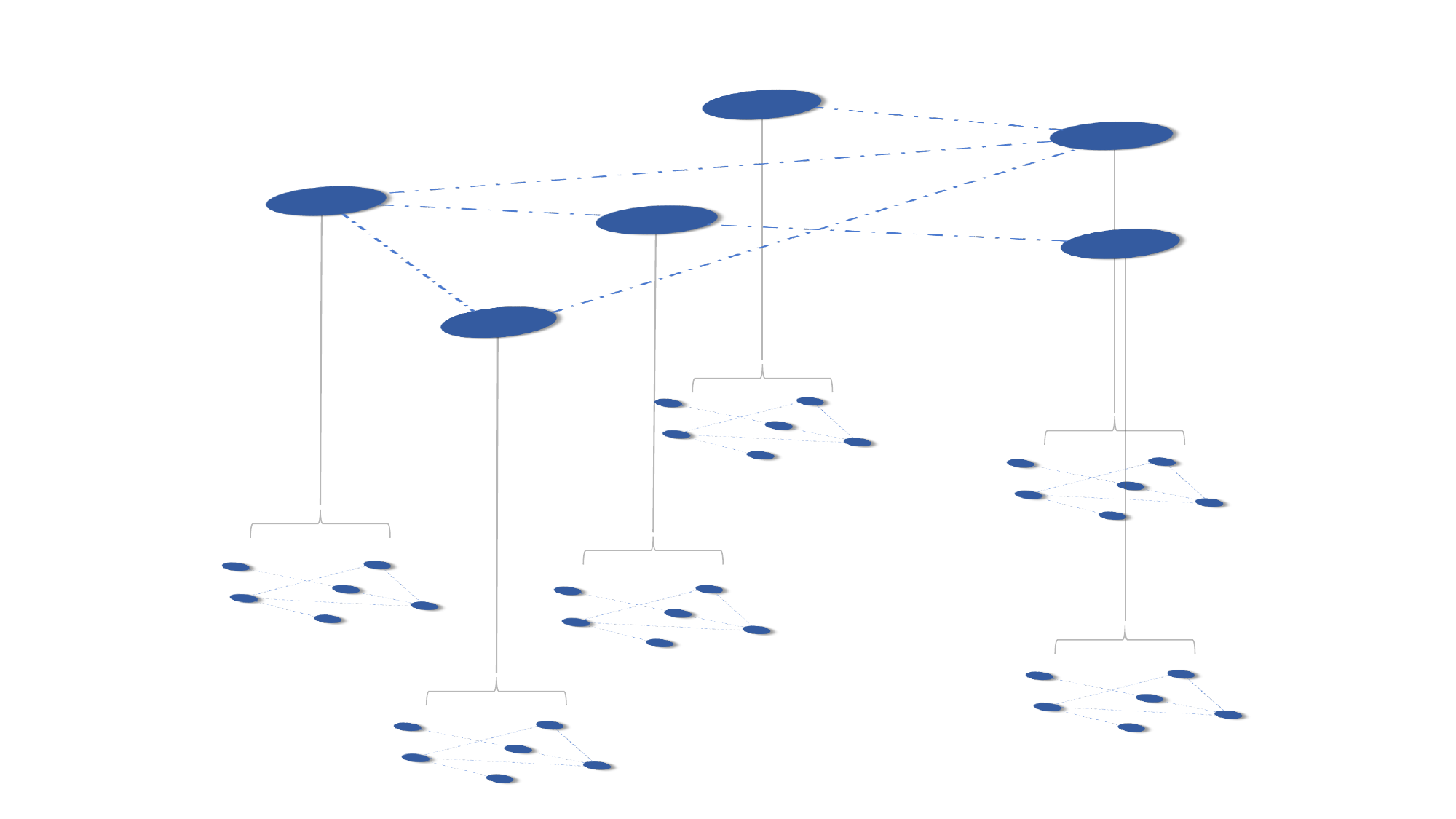}
\caption[h!]{\label{fig:Clusteringpicture} Clustering protocol illustration. Nodes with similar  connectivity in the requests are grouped in clusters, where each of them is identified by a router or representative node in a higher layer, where clustering algorithm can be iteratively applied again. Resource states are appropriately designed for each layer.}
\end{figure}

We introduce a spectral clustering protocol that  exploits the heterogeneity of the connections of the requests. Among the possible target configurations, one can find links that are very unlikely to appear, or regions of strong connectivity. We propose to make use of clustering techniques from graph theory that output a hierarchical structure that leads to improved storage and entanglement requirements in the  resource state. The functionality in each cluster (or subnetwork) is guaranteed by a specific multipartite resource state, and this cluster is then replaced by a single router node at the next layer. Connections between router nodes at the next layer are provided by another network resource state (see Fig. \ref{fig:Clusteringpicture}). The basic underlying idea why this leads to a reduction of the required resources relies on the fact that connections between nodes in different clusters are sparser than within a cluster. It is hence sufficient to provide a single resource state between router nodes of the clusters, which suffices to provide all the required connections between clusters. A typical situation where one benefits from such an approach are network requests with multiple parallel connections within clusters, but only few connections between different clusters in all desired target configurations.

In particular, we apply a spectral clustering algorithm based on the distribution of links in the request configurations. Given the definition of the  graph laplacian introduced above (see Eq. (\ref{eq:laplacian})), the spectral clustering we propose here involves the following steps:

\begin{enumerate}
\item Construct the cumulative matrix $C$ (Eq. \ref{eq:cumulativematrix}) based on all the possible requests and compute the corresponding laplacian operator $L$  (Eq. \ref{eq:laplacian}).
\item Obtain the eigendecomposition of $L$ and its $k$ largest eigenvectors. The parameter $k$ is chosen in an heuristic way identifying the largest difference between consecutive eigenvalues. 
\item Apply $k-$means clustering \cite{vonLuxburg2007,Likas2003} to group the network nodes into $k$ different clusters.
\end{enumerate}

Once the different clusters are defined, each of them can be identified  with a new single node in a higher network layer. This router node represents all nodes in the subnetwork or cluster, and guarantees connections of all these nodes to the remaining network. 

Connections between the clusters are recomputed from the adjacency matrix and the requests, such that the clustering protocol can be iteratively applied again for each higher layer, leading to a hierarchical structure.

{\LinesNumberedHidden
    \begin{algorithm}
        \SetKwInOut{Input}{Input}
        \SetKwInOut{Output}{Output}
        \SetAlgorithmName{Algorithm}{}

       \justifying \textit{Input}: $m$ different  request configurations.
        
        \begin{enumerate}
            \item Compute the cumulative matrix $C$ (Eq. \ref{eq:cumulativematrix}) and the corresponding laplacian $L$ (Eq. \ref{eq:laplacian}).
            \item Find the $k$ largest eigenvalues of $L$ and apply a $k-$means clustering algorithm, that identifies groups of $k$ elements.
            \item Consider each cluster as a single node in a higher layer, recompute $C$ and $L$ and repeat iteratively.
        \end{enumerate}
        
     \justifying \textit{Output}: Multi-layer hierarchical structure with nodes grouped by their cumulative connectivity in the different requests.
\caption{Quantum network clustering}
\end{algorithm}}

Clustering allows us to group nodes into smaller groups with homogeneous connectivity. The resources are built independently for smaller and more homogeneous groups of nodes. This leads to groups of entangled-based resource states comprising less number of qubits and less complex structures, with reduced memory requirement. In addition, entanglement of the resource network states is therefore easier to establish (during the dynamic phase) and to maintain (static phase).

It is important to stress that the clusters identified by this approach do not correspond to the underlying physical topology of the network, i.e. the channels that connect different nodes. Components of a cluster may rather be distributed over the whole network without any direct physical connection, and the algorithm outputs an improved way to group the nodes, such that the overall storage requirements are minimized. Hence, we introduce an optimized topology to the network state that solely depends on the required functionality, where links are provided by shared entanglement between nodes. This is a unique possibility that arises from the use of entanglement as a tool to realize quantum communication, where the usage of multipartite entangled states allows one to further reduce resources.

We note that the clustering algorithm would work in the same way in case the request configurations would include also multipartite connections.  

\subsection{Full pairwise connectivity approaches}\label{sec:Full connectivity protocols}
We now consider a situation of particular interest, i.e. when the resource state has to guarantee full pairwise connectivity.  Target configurations in this case include any combination of simultaneous, bipartite pairwise connections (i.e. a collection of Bell states). Observe that for a network of $n$ nodes each request can contain $\frac{n}{2}$ parallel bipartite links at most. If all the possible combinations of $\frac{n}{2}$ Bell-type connections can be guaranteed from some network state (full pairwise connectivity), any request with $< \frac{n}{2}$ can also be obtained. We therefore look at network states that can be transformed into any of the possible request with $\frac{n}{2}$ links (see Appendix \ref{Sec:Appendix2}). Note however that the number of these different requests scales as $(n-1)!!=\prod_{k=1}^{n/2} (2k-1)$. In case only some of these configurations are demanded by the network, better strategies may exist (see Sec. \ref{sec:merging}).

Observe that we allow for target connections involving any pair of nodes of the network here. Many realistic situations however involve only connections between --and not within-- two sets of nodes. The so-called butterfly state is an example of this kind. Two input nodes can be connected to two output nodes in two different configurations, which can be accomplished by a six-qubit resource state and allows one to resolve network conjunctions \cite{Hahn2019,Satoh16}. For sake of clarity, we analyze this case in Appendix \ref{Sec:Appendix4}. 

\subsubsection{Bell-type construction}
A trivial solution for full pairwise connectivity consists in simply providing a tensor product of all possible Bell pairs between the nodes of the network, as shown in Fig. (\ref{fig:Fullconec_picture}.a). In order to achieve full pairwise connectivity, one requires $2\sum_{i=1}^{n} (n-i)=n(n-1)$ total number of qubits stored by the $n$ nodes of the network. We recall that we are considering a top-down network approach, where the network entangled state is provided --and stored-- before each request is demanded. Therefore, providing all possible Bell states, and keeping in each request the ones demanded, is not the best strategy in general. We refer to Sec. \ref{Sec:performanceclustering} for an analysis under different assumptions.

\subsubsection{Switch-type construction}
An alternative solution based also in Bell states is the one shown in Fig. (\ref{fig:Fullconec_picture}.b), closely related to entangled switch constructions \cite{Vardoyan2019,Vardoyan2020,Coopmans2021}. This approach, which can be proven to be optimal (see Appendix \ref{Sec:Appendix2}) for $4-$nodes networks in terms of storage, can be generalized for an arbitrary network of $n$ nodes, with an storage scaling of $ 2(n-1)$, that improves the direct Bell-type approach in a factor $n$. 

The construction however requires that one single node stores many of the qubits ($n-1$), and multi-qubit (although local) operations are required in that node to guarantee any request. In Appendix \ref{Sec:Appendix2} we provide illustrative details of how each request can be guaranteed. The network therefore experiences a strong dependence on that node and the stored qubits are distributed among the network nodes in a much less uniform way than other strategies (and hence more sensitive to losses of certain nodes), specially for large networks. 

\subsubsection{GHZ-type construction} \label{sec:GHZ-type construction}
\begin{figure}
\includegraphics[width=\columnwidth]{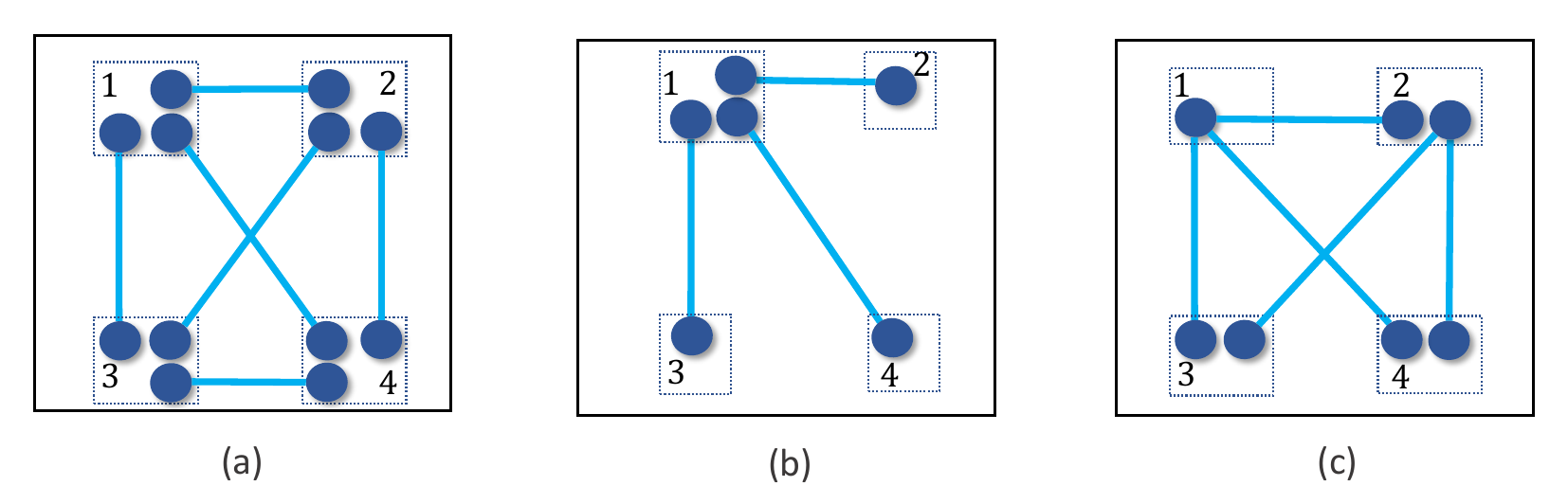}
\caption[h!]{\label{fig:Fullconec_picture} Network resource states that guarantee any request involving --simultaneous-- bipartite connections in a $4-$node network. Three requests have to be guaranteed, i.e. $\{1$-$2$, $3$-$4\}$, $\{1$-$3$, $2$-$4\}$ and $\{1$-$4$, $2$-$3\}$. See also Appendix \ref{Sec:Appendix2}. (a) Bell-type direct construction ($12$ qubits). (b) Switch-type construction ($6$ qubits), which is the optimal solution. (c) GHZ-type construction ($7$ qubits).}
\end{figure}

An different solution is based on multipartite entanglement, more concretely, on the usage of GHZ states \cite{Pirker2018,Pirker2019}. It is easy to see that, in order to guarantee the desired full pairwise functionality in a network of $n$ nodes, one needs $n/2$ different GHZ states of progressive reduced size, i.e. $\sum_{i=0}^{n/2-1} (n-i)=\frac{1}{8}n(3n+2)$ total number of qubits (see Fig. (\ref{fig:Fullconec_picture}.c). This is sufficient to allow for all the possible $n/2$ parallel pairwise connections, i.e. all target configuration we consider. Although this construction requires more storage than the switch-type one, all the requests can be achieved with single-qubit operations, and the network does not rely on a single node storing most of the qubits. 

\subsection{Merging protocol} \label{sec:merging}
Whenever only some target configurations have to be guaranteed, solutions introduced in the previous section can be very costly. In such situations, a direct construction of the resource state taking the specific requests into account is beneficial. We introduce here a novel algorithm that iteratively construct tree-like entangled resource states by exploiting the simultaneity of the connections, i.e. how often certain links appear at the same time in the different requests. This leads to a significant storage enhancement.  Recall that, as stated in Sec. \ref{Sec:Optimality}, deciding whether a graph state can be transformed into another one is in general a NP-Complete problem, and therefore, one cannot expect to design a protocol that achieves optimal solutions for any situation. However, as we demonstrate below our approach provides a significant memory reduction, which in some situations can be shown to be optimal or close to optimal. 
The idea behind the merging protocol can be understood with a simple example. Imagine one node of the network $a$ which in one request needs to be connected with some node $b$, while in other request it needs to be connected with some node $c$. The merging algorithm first virtually builds two Bell pairs $\{a$,$b\}$, $\{a$,$c\}$. Then, it takes both qubits of $a$ and decides whether they can be merged into a single one. In this case, since the two connections appear in different requests, the qubits are merged, therefore generating a GHZ between $\{a,b,c\}$. This GHZ state can be transformed to either a Bell state shared between $ab$, or $ac$, and guarantees the desired functionality. 

\begin{figure}
\includegraphics[scale=0.25]{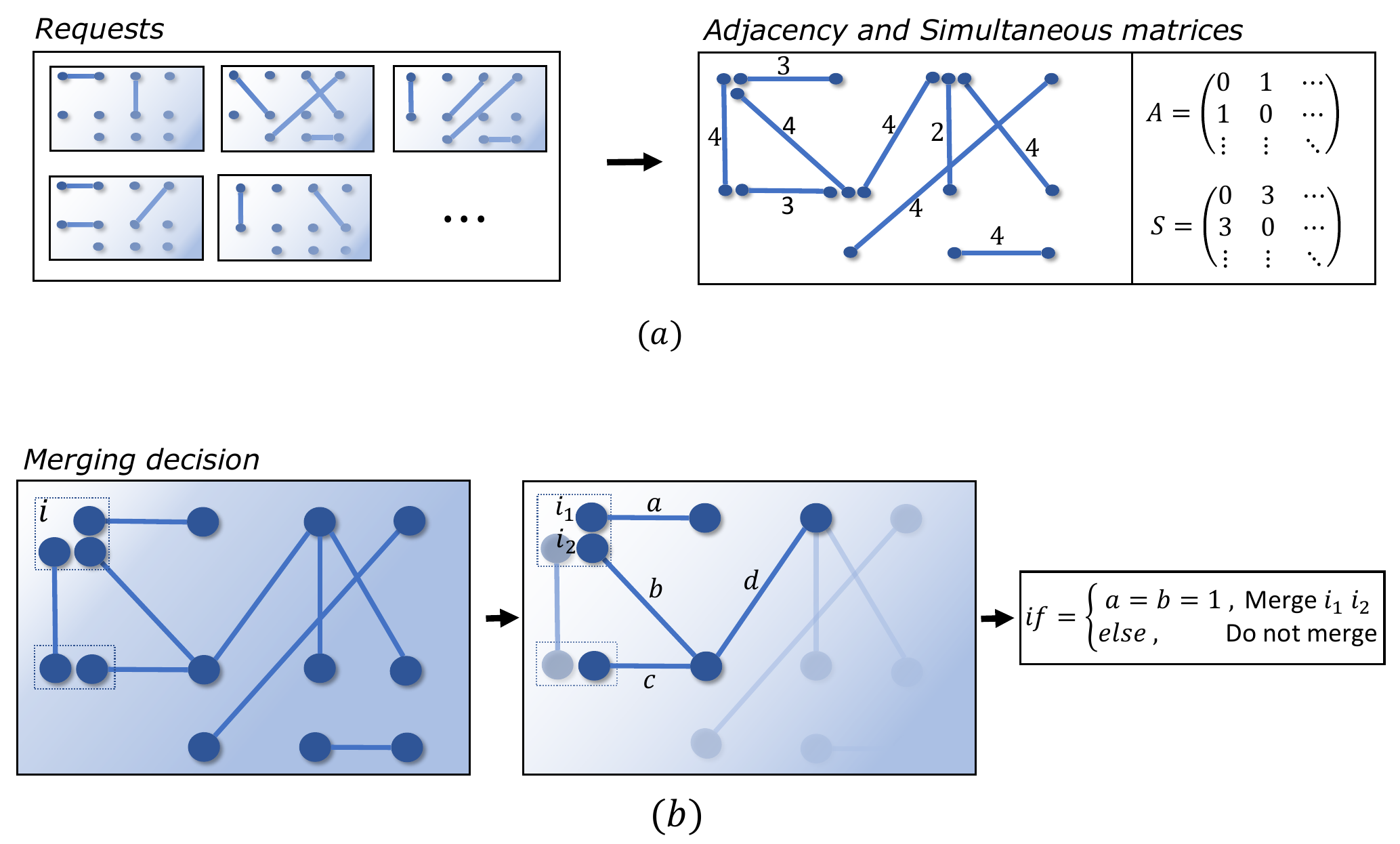}
\caption{\label{fig:Mergingpicture} Merging algorithm illustration. (a) Initial scheme. From all the requests that have to be guaranteed (left), the adjacency (Eq. \ref{eq:adjmatrix}) and simultaneous (Eq. \ref{eq:simultaneousmatrix}) matrices are computed based of the simultaneity of each connection in the requests (Eq. \ref{eq:simultaneousmatrix}) (b) Certain step of the merging process. Node $i$ decides whether two of its qubits, $i_{1}, i_{2}$, are merged. For that purpose, only the $1^{st}$ and $2^{nd}$ neighborhoods of the qubits are considered. The simultaneous matrix is recomputed for this subspace and the qubits are merged depending on the incident-edge weights of $i_{1}$ and  $i_{2}$ in this matrix, i.e. $a,b$. Note that the qubits and merging operations do not have to be physically implemented, and the resource state is only physically built once the merging algorithm is applied.}
\end{figure}

We consider a network with $n$ nodes and $m$ different requests that the network resource state should guarantee. This can certainly be guaranteed by considering the union of all states appearing in all the requests, which we take as a starting point. Notice that states that appear in multiple requests need to be considered only once. The merging algorithm then aims at combining some of these states to multipartite states in such a way that all desired target states can still be reached by local operations. The merging algorithm comprises the following steps:

\begin{enumerate}[label=(\roman*), leftmargin=\parindent,align=left,labelwidth=\parindent,labelsep=2pt]
\itemsep0em
\item Initially, the  adjacency matrix $A$ (see Sec. \ref{Sec:Setting details}) is computed based on the $m$ target requests. Bell states are virtually built accordingly for each non-zero entry of the matrix $A$, such that several --virtual-- qubits are stored per node.  We remark that the Bell states do not have to be physically constructed. The resource state is only built after the merging process is complete. We therefore refer to these qubits as virtual qubits for the moment.

\item The simultaneous matrix (Eq. \ref{eq:simultaneousmatrix}) is computed. For each node $i=\{1,...,n\}$,  virtual qubits are considered two-by-two. Starting from the two first prepared virtual qubits $i_{1}, i_{2}$ of node $i$, a decision of whether the two qubits are --virtually-- merged into a single qubit is made based on their incident-edge weights in the following way (see Fig. \ref{fig:Mergingpicture}): 

  \begin{enumerate}[label=(\arabic*)]
    \item  Consider the incident edges of qubits $i_{1}$ and $i_{2}$. If the maximum value over all the incident edges weights of the simultaneous matrix is $\leq 1$, the two qubits are (virtually) merged, i.e.
    \begin{equation}
    \mathrm{if}\,\,\,\max_{j}S_{i_{1},j}\leq1\,\,\,\&\,\,\,\max_{j}S_{i_{2},j}\leq1\,\,\,\rightarrow\,\,\mathrm{merge}
    \label{eq:merge1}
    \end{equation}
    If condition Eq. (\ref{eq:merge1}) is not satisfied, move to the next step. If condition Eq. (\ref{eq:merge1}) is satisfied, qubits $i_{1}, i_{2}$ are merged (via the operation Eq.(\ref{eq:merging})) into one $i_{1}'$, and the process is started again.
    
    \item Compute the simultaneous matrix $S'$ (Eq. \ref{eq:simultaneousmatrix}) but restricted to the reduced set corresponding to the first and second neighborhood of $i_{1}$ and  $i_{2}$, i.e. with elements $S'_{i,j}$ for \small $\{i,j\} \in \{ i,j,N(i),N(j),N(N(i)),N(N(j)) \}$\normalsize.  Observe that not only the weights of the edges outside this set equals $0$ in $S'$, but also the weights of the remaining entries change in general. 
    
    \item Repeat step $(1)$ but with the simultaneous matrix $S'$. If it succeeds, merge the qubits and start again. Otherwise, move to the next step.
    \end{enumerate}
   \item If both previous steps fail for a pair of qubits, they are not merged. Choose the next combination of qubits, i.e. $i_{1}, i_{3}$, and proceed again.
  \end{enumerate}


The protocol iteratively merges qubits in a virtual way, generating larger multipartite entangled states. Importantly, the output state of the protocol only requires single-qubit operations to guarantee any request. The algorithm could be further improved by considering all the possible local-complementation orbits \cite{Hein2006} of each state before each merging decision. We however have not implemented this option in our simulations.

{\LinesNumberedHidden
    \begin{algorithm}
        \SetKwInOut{Input}{Input}
        \SetKwInOut{Output}{Output}
        \SetAlgorithmName{Algorithm}{}

        \textit{Input}: $m$ different request configurations.
        
        \begin{enumerate}
            \item Compute the adjacency matrix $A$ (Eq. \ref{eq:adjmatrix}) and virtually consider all the bipartite links.
            \item Take the virtual qubits of each node $i$ two by two $i_1,i_2$. Compute the reduced simultaneous  matrix $S'$ (Eq. \ref{eq:simultaneousmatrix}) corresponding to the $1^{st}$ and $2^{nd}$ neighborhoods of $i_1$ and $i_2$. 
            \item If the corresponding weight of \textit{all} the incident vertices to the two qubits $i_1,i_2$ is $S'_{ij} \leq 1$, they are merged.
            \item Repeat for all qubits and every node and construct the final state.
        \end{enumerate} 
        
        \justifying  \textit{Output}: Graph-type entangled network state with enhanced storage that can guarantee any of the $m$ input requests by single-qubit operations.
\caption{Quantum network merging}
\end{algorithm}}

Once the protocol finishes and all the merging decisions have been considered, the final network state is designed and can be physically generated. This final state exhibits a tree-like structure that significantly reduce the total memory requirements. As an example, consider the case each of the request configurations only involve one Bell connection between any two nodes. Independently of the number of requests, the optimal resource state consists of a single GHZ state of $k$ particles, where $k$ is the number of particles involved in the requests, since a $k-$party GHZ state can be transformed into a Bell state between any two constituents (see Sec. \ref{sec:basicops}). Since only one qubit has to be stored per node, optimality in this case is guaranteed. The merging algorithm output in this case is exactly this $k-$qubit GHZ state, since all merging decisions succeed. 

Although we analyze in the following the performance of the merging protocol under the same assumptions introduced before, i.e. each request can only contain bipartite connections and each node can be part of only a single connection (i.e. be connected to one other node at most), we remark that the merging algorithm can be directly adapted and applied in more general situations, where target configurations involve arbitrary multipartite graph states shared by certain network nodes, or for cases where the nodes are connected to several other nodes in each request. When these assumptions are relaxed, other approaches with similar storage performance (e.g. switch-type construction) are not longer applicable.

The merging protocol, as well as the full pairwise connectivity approaches, can be included in combination with the clustering procedure:  Once the clustering algorithm is implemented, one can apply the merging algorithm --or provide the full-connectivity solutions-- for each of the individual clusters generated in each layer, in order to further enhance the storage requirements of the network state.

Both merging and clustering strategies can be generalized for situations where the requests also involve multipartite arbitrary connections. A direct solution consists in defining the simultaneous matrix (Eq. \ref{eq:simultaneousmatrix}) in such a way that all the elements belonging to the same connected graph only contribute once to $s^{q}_{i,j}=\sum_{l,r}A_{l,r}^{q}$. This allow the protocols to work without additional overheads or complexity. However better solutions may exist.

{\LinesNumberedHidden
    \begin{algorithm}
        \SetKwInOut{Input}{Input}
        \SetKwInOut{Output}{Output}
        \SetAlgorithmName{Algorithm}{}

        \textit{Input}: $m$ request configurations.
        
        \begin{enumerate}
            \item Apply iteratively $s$ rounds of the clustering algorithm such that nodes are grouped in a multi-layer structure.
            \item Apply the merging algorithm in each of the clusters of each of the $s$ layers independently.
        \end{enumerate} 
        
        \justifying \textit{Output}: Multi-layer resource state that can guarantee the $m$ requests with single-qubit operations. The resulting multipartite entangled resource state exploits the benefits of clustering and merging algorithms, with optimized storage and entanglement requirements.
\caption{Combined clustering and merging}
\end{algorithm}}

\section{Results, performance and examples}
In this section we analyze the performance of the different protocols --and combinations of them-- introduced before. Given the nature of the problem, our results mainly involve numerical studies. On the one hand, we analyze cases where the possible requests are generated in a random way. Additionally, we examine situations where the optimal solution is known, in order to benchmark the performance of the different approaches. 

We recall that we analyze the deterministic scenario. In this static situation, the task is to find the optimal entangled-based resource state that can guarantee or fulfill any single of the requests deterministically with local operations. We consider other scenarios in Sec. \ref{sec:Probabilistic scenario}.

\subsection{Performance of the merging algorithm}
We start by studying the performance of the merging protocol under different initial conditions. Consider a network of $n$ nodes where the entangled-based resource state has to guarantee $m$ different requests generated in a random way, consisting of bipartite links. Each node has the same probability of sharing a connection with another one, but only one connection per node is allowed per request. Note again that, in each target configuration, at most $\frac{n}{2}$ simultaneous links can appear under these restrictions. 

In order to properly interpret the results, let us first consider the extreme cases. First, assume that each of the $m$ target configurations involves just one link between some two nodes, and each link appears in one target configuration. Therefore, the optimal resource state associated is simply a connected $n-$qubit graph state, e.g a $n-$party GHZ state, where each qubit of the GHZ state belongs to each node of the network. The total memory required scales then linearly with the number of users $n$, i.e. each node has to store a single qubit only. In contrast, consider now that the target requests are given in such a manner that all possible combinations of --simultaneous-- bipartite links are taken into account. In this case, a full pairwise connectivity approach (see Sec. \ref{sec:Full connectivity protocols}) offers the best solution. The storage scaling for these constructions ranges from linear (e.g. switch type construction) to quadratic (e.g. GHZ approach) with the number of nodes. Notice that a larger than quadratic overhead might be required if one also considers multipartite entangled states as possible target states, or in situations where each node is connected to multiple other nodes -- and not just a single one as we consider here. 

Observe that when we refer to a linear storage scaling, this corresponds to a constant scaling in the number of qubits stored per network node.  

\begin{figure*}
 \centering
\includegraphics[scale=0.5]{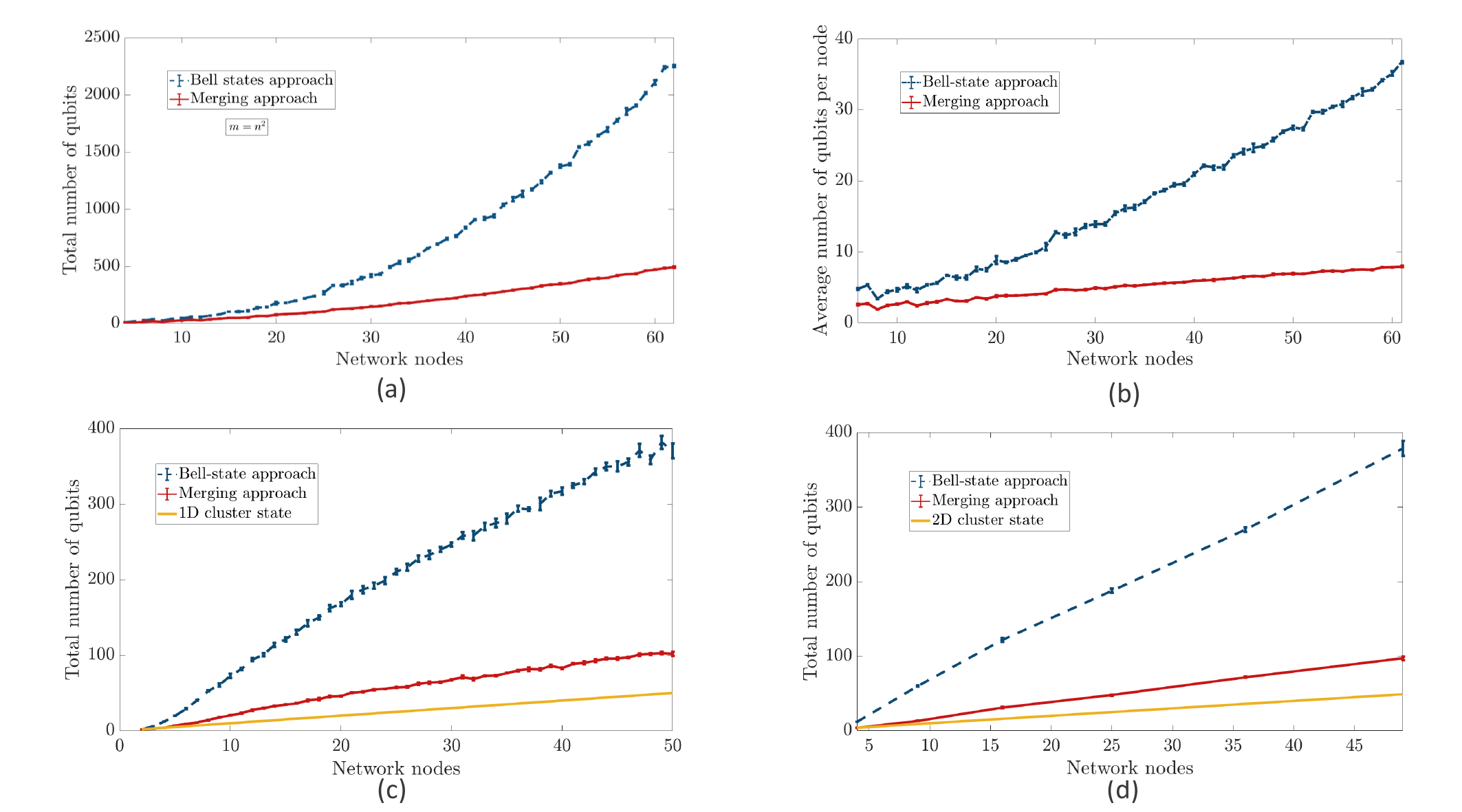}
\caption{\label{fig:onlymerging} Performance of the merging algorithm. (a) and (b) shows the total storage and averaged storage per node, associated with the merging algorithm with $m=n^2$ and $m=2n$ randomly generated request states respectively, for different network sizes ($n$). (c) and (d) shows the performance of the merging algorithm where the requests correspond to states that can be reached from the 1D and 2D cluster states.  See Appendix \ref{Sec:Numerics} for details. }
\end{figure*}

Fig. (\ref{fig:onlymerging}.a) and (\ref{fig:onlymerging}.b) show different numerical results for networks of $n$ nodes and $m$ randomly generated request configurations that can contain simultaneous links (see Appendix \ref{Sec:Numerics} for details). The network resource state, i.e. the state that the algorithm outputs, is able to guarantee any of the $m$ requests with single-qubit operations.  We consider different parameters which represent intermediate scenarios between the extreme cases explained above. 
Note first that, because of the NP-Complete nature of the problem, the optimal solutions cannot be obtained efficiently, and therefore we compare our merging approach with respect to the trivial case of providing the minimum number of Bell pairs. One can observe that the merging algorithm performs significantly better in terms of storage. More concretely, numerical fitting shows that the merging approach exhibits a quasi-linear scaling (observe in Fig. (\ref{fig:onlymerging}.b) the almost constant scaling in the number of qubits per node), which indicates that the algorithm output should be close to the hypothetical optimal strategy.

We can also perform an alternative analysis consisting in choosing some specific resource state and considering all the target requests that can be reached from it. One can afterwards provide these requests as inputs for the merging algorithm to check how well it behaves with respect to the chosen, and hence optimal, resources. Ideally, it would reconstruct the initial resource state where only a single qubit per node is stored. In particular, we analyze this scenario for the 1D and 2D cluster states, for which we know what kind of configurations can be reached. 

Cluster states are graph states that correspond to a specific lattice structure. They allow one to establish entangled pairs between any two parties. For instance, a 1D cluster state of size $3n-1$ allows one to prepare $n$ Bell pairs between nearest neighbors by measuring every third qubit in the $Z$-basis. The distance of links can be increased by measuring the intermediate qubits in the $Y$-basis, thereby reducing the number pairs that can be established in parallel. Similarly, in a 2D cluster state, measuring the six neighbors of two neighboring qubits generates a Bell state that is decoupled from the rest of the structure, which is still intact and can be used to generate other Bell pairs. Note that, for all the target configurations that can be reached from the 1D and 2D cluster state, a direct resource construction based on Bell pairs would require a number of qubits that scales quadratic in $n$, whereas this scaling is linear for the cluster states. We consider all possible configurations that can be reached by --a combination of-- repeater path protocols (see Sec. \ref{sec:basicops}). We refer to Appendix \ref{Sec:Numerics} for more details.

These target configurations are provided as inputs for the merging protocol. In case we restrict to only nearest-neighbor connection, the merging algorithm exactly outputs the 1D and 2D cluster states respectively. If one considers all the possible requests, Fig (\ref{fig:onlymerging}.c) and (\ref{fig:onlymerging}.d) shows how merging protocol performs as a function of the number of network nodes. One can observe that the merging algorithm performance is close to the optimal resource states, i.e. the 1D and 2D cluster states.

\subsection{Performance of the clustering algorithm and the combined protocols} \label{Sec:performanceclustering}
The clustering protocol takes advantage of the heterogeneity of the network connectivity in order to structure the network in a multi-layer way, leading to benefits in terms of storage and entanglement. Although we show in this section performance analyses concerning storage advantages, we remark that the multi-layer architecture provided by the clustering protocol also leads to significant reduction in the complexity of the entanglement structure of the resource states, which implies benefits in establishing and maintaining these states. 

\begin{figure*}
 \centering
\includegraphics[scale=0.5]{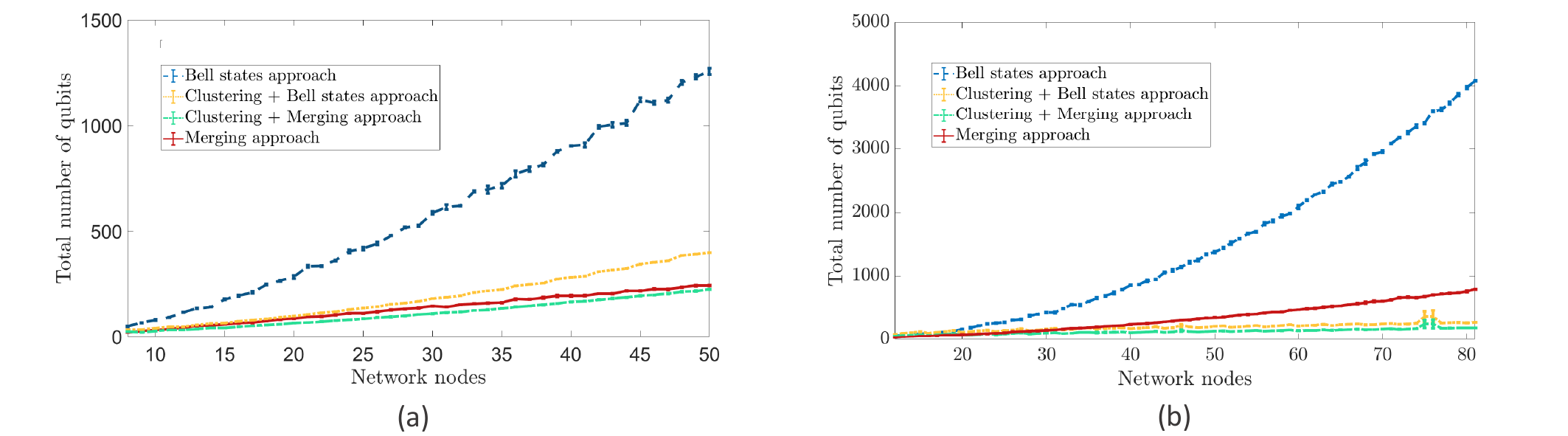}
\caption{\label{fig:Clusteringfigures} Performance of the clustering (dotted yellow) and the clustering+merging (dashed-dotted green) algorithm, under different conditions and compared with directly providing the necessary Bell pairs (dashed blue) or only applying the merging algorithm (solid red). Requests are taken randomly but with modified probability to cause heterogeneity in the connections (see Appendix \ref{Sec:Numerics} for details). (a) Average number of qubits per node for $m=n$ random different requests and higher probability for connections among groups of $4$ nodes. (b) Several rounds of clustering with $m=2n$ requests and higher probability for connections among groups of $4$ nodes.}
\end{figure*}

Fig. \ref{fig:Clusteringfigures} shows several results for different initial settings. For each point, a network of $n$ nodes is considered and $m$ different requests are randomly generated. Heterogeneity in the connections is forced by giving more probability to some connections when two nodes belong to a same pre-defined group (see Appendix \ref{Sec:Numerics}). The requests are provided as inputs for the clustering algorithm, which is applied once (Fig. \ref{fig:Clusteringfigures}.a) or twice in an iterative way (Fig. \ref{fig:Clusteringfigures}.b). Once the clustering algorithm groups the network nodes according to the connectivity in the different requests, each cluster is treated independently, applying full pairwise connectivity solutions (Sec. \ref{sec:Full connectivity protocols}) or implementing the merging algorithm for each cluster. The output state is such that it can be transformed into any of the $m$ requests by local operations. 

The advantage provided by the clustering process itself is significant. Nevertheless, when it is combined with the merging algorithm applied in each cluster, the storage benefit can be further improved. The combined protocol exhibits a quasi-linear scaling, as one can see from the almost constant behaviour in the total number of qubits required (Fig \ref{fig:Clusteringfigures}.b).  We refer to Appendix \ref{Sec:Numerics} for details. 

Although the merging algorithm itself sometimes performs similarly than the combined algorithms, the entanglement structures of the network states generated are much simpler in the combined case, and thus the entanglement is easier to establish and maintain in realistic situations.

\section{Probabilistic scenario} \label{sec:Probabilistic scenario}
We have analyzed a deterministic setting so far. There a resource entangled state is optimized in such a way that any of $m$ given requests can be guaranteed via local transformations of the resource. In other words, it is assumed that after each request is completed, the resource state (or at least the part that has been consumed) is built and provided again.

In this section we consider an alternative scenario. Considering the same $i=\{1,\dots,m\}$ requests as before, each of one with certain probability $p_i$ with $\sum_{i=1}^{m} p_i=1$, the task now is to complete, up to some failure probability, $k$ request calling rounds with a single resource state. A trivial solution, that succeeds with unit probability, consists in providing $k$ copies of the resource state obtained in the deterministic case, however better solutions may exist. 

This can be seen with a simple example. Consider three nodes $a,b,c$ and two possible requests, one involving the connection $(a,b)$ and the other involving $(a,c)$, each of one with equal probability for each call. In case $k \gg 1$ requests has to be satisfy, the deterministic approach tells us that $k$ GHZ states between the three parties suffice, with $3k$ qubits in total. However, in the $k$ calls we will find on average half $(a,b)$ and half  $(a,c)$ connections, and therefore building just the corresponding Bell pairs ($2k$ qubits in total) is enough to fulfill the task, up to some failure probability arising from statistical fluctuations. 
In this section we discuss several scenarios where certain thresholds can be found and where we assume some failure probability is accepted. A detailed analysis of this problem lies however beyond the scope of this work.

\subsection{Single round}
When only a single request run, $k=1$,  has to be performed, the situation hence reduces to the deterministic case treated before. However, when allowing for a small probability of failure, some requests with a low probability of occurrence can be neglected and the deterministic case can output much beneficial states in terms of storage. Notice that the choice is not unique, as several requests might be ignored, as long as the overall probability of them to occur is below the allowed failure probability, and depending on that choice the required storage resources may differ significantly. 

\subsection{Asymptotic limit of multiple rounds}
In the asymptotic limit of request calls, $k \to \infty$, the optimal solution is to provide exactly the states corresponding to the $m$ target configurations $\{(p_j,|\psi_{G_j}\rangle)\}$, where the number of copies depends on the probabilities, $|\Psi\rangle=\otimes_{j}^{m} |\psi_{G_j}\rangle^{\otimes p_j k}$. This corresponds to the minimal number of stored qubits, as no manipulation of states is required, and exactly the states that are requested are consumed. 

In case requests only involve Bell pairs shared between multiple pairs of nodes (as we considered in the deterministic scenario), the optimal solution corresponds to storing the Bell states directly, where the required number of copies of a certain Bell pair is given by the total number it appears in any of the requests. For a Bell pair between nodes $\mu,\nu$, we have that the required number of copies is given by 
\begin{equation}
\label{Bellstatesasympt}
n_{\mu\nu}=\sum_{\{j|(\mu,\nu) \in E_j\}} p_j k, 
\end{equation}where $E_j$ is the edge set corresponding to graph $G_j$. That is, if the Bell pair is part of the request $j$, we need to store a copy, which occurs with probability $p_j$. It follows that the total number of qubits to be stored is upper bounded by $k n$, which is only reached if all requests contain $n/2$ parallel Bell pairs. Notice the linear scaling in the required resources.

\subsection{Finite number of rounds}
What is more interesting is an intermediate regime of a finite, small number of requests calling rounds $k$. There, we can make use of both, the advantage resulting from clustering and merging as in the the deterministic approach, as well as the fact that some states are required with smaller probabilities and hence fewer copies need to be stored.  In the remainder of this section we describe different examples of possible approaches.

\subsubsection{Single Bell pair requests}
To illustrate this, consider first the case of all possible combinations of requests consisting of a single Bell pair, 
each occurring with probability $p_{j}=[n(n-1)/2]^{-1}$. We know that a single GHZ state suffices to fulfill any of the requests, so $k$ GHZ states, i.e. storing $nk$ qubits in total, allows to fulfill $k$ requests. In turn, each Bell pair occurs on average $kp_j$ times, and hence it suffices to store $2k$ qubits on average. However, this consideration only is applicable if $kp_j>1$, i.e. each Bell pair occurs on average at least once in the $k$ requests. Otherwise, one needs to store $\lceil kp_j \rceil$ copies of each Bell state, i.e. at least one copy of each Bell state.
As long as $k \leq (n-1)/2$, the GHZ approach is then favourable. 

\subsubsection{Simultaneous Bell pair requests - weight separation}
Consider a more general scenario, where there exist $m$ possible target configurations containing bipartite (simultaneous) connections, and where each node can only be connected with a single other node in each request.  Assuming for simplicity that all the request configurations have the same probability, one can assign each of the links a weight corresponding to the number of requests where that link appears over $m$. A layer separation is hence built attending to these weights, such that each layer comprises edges whose weights fall into the same interval. Each layer is now treated independently, following one of the next approaches.

The most direct strategy consists in providing Bell pairs with certain probability $p$. If $k$ requests call rounds have to be fulfilled, then a certain Bell pair has to be provided $pk$ times, where $p$ is the weight of the link, i.e. $p=\frac{q}{m}$, with $q$ the number of requests the link appears. Observe that if $pk \leq 1$ at least one Bell state should be provided. However, there exists a threshold from where merging --or multipartite approaches-- is advantageous with respect to this direct Bell approach. The threshold lies in the $kp \leq 1$ interval and varies with the setting and the output of the merging approach. One example has been stressed above, where each request involves only one Bell pair and all the variants are possible, and the regime from which a GHZ approach is favourable is $k \leq (n-1)/2$.

\subsubsection{Combination with merging and clustering}
Consider now a cluster state of size $n$ and requests consisting of all possible combinations of $n/2$ simultaneous links, where each of the simultaneous connections appears with probability $p_i$. The total cluster weight is then simply $p_c=\sum p_i$. In this scenario, the merging algorithm (Sec. \ref{sec:merging}) outputs a GHZ full-connectivity state of $s=\frac{n}{8}(3n+2)$ qubits (see Sec. \ref{sec:GHZ-type construction}). One should then provide $s k p_c$ qubits for fulfilling $k$ requests, up to some fixed failure probability. Using the direct Bell approach one requires $2 k \sum p_j$ qubits for the $j$ different single links, such that in the regime where each $k p_j \leq 1$, one need to provide at least $n(n-1)$ qubits. Observe that $p_j$ refers to probabilities of single links whereas $p_i$ refers to probabilities of simultaneous requests, i.e. $n/2$ links per request.  The merging strategy becomes then advantageous in the  $k p_c \leq \frac{8(n-1)}{3n+2}$ regime. For instance,  if $n=4$,  where the three possible cluster requests are possible, the merging approach is advantageous if $k p_c \leq 1.71$, or equivalently if $k p_j \leq 0.57$ for any $j$. 

One could additional include the spectral clustering strategy (Sec. \ref{sec:clusteriing})  in each of the weighed layers, and treat each cluster in the following way. First, the absolute weight of the cluster is defined as $p_c=\frac{q_c}{m}$, where $q_c$ defines the number of target configurations any of the cluster links appear. The merging algorithm is applied afterwards, taking into account that one should provide $k p_c$ copies of the resource output state. If this result outperforms the direct Bell approach in the cluster, then the merging is kept. 

Note that, if there exist requests where not all the possible $n/2$ simultaneous configurations appear, their contribution to $p_c$  is also weighed. For instance, in a $4-$node cluster, requests with only one connection contributes with a factor $1/2$ to $p_c$, since one merged state can guarantee two of these connections. 

\subsubsection{Clustering based on probabilities}
We finally discuss another alternative approach for the case where each request has different probabilities assigned. It consists of an additional clustering of the target configurations based on the probabilities they occur. That is, we define sets of target configurations that have similar probability of occurrence, with the aim to identify sets where further reduction of resources is possible. For instance, we can define set $S_m$ to contain all request configurations for which the probability $p_j$ fulfills  $10^{-m} \leq p_j \leq 10^{-m+1}$. We can then treat each of the sets independently, and determine if a resource reduction using any of the methods discussed above (or a combination thereof) is possible. This includes clustering or state merging, as well as a possible reduction based on using only the average number of required resource states if some states/configurations appear on average more than once when we consider $k$ requests. For small values of $m$, the probability for target configurations is large, and hence multiple copies of certain states are required in case of small number $k$ of callss. In turn, if $m$ is large, probabilities are small and hence on average only a few (or less than one) state is needed. In this case, it is likely that one cannot rely on producing copies of Bell states only with respect to their relative occurrence, but one may be able to gain by using clustering and state merging as in the deterministic approach, where the reduction based on multiple repetitions might become available as the overall effective probability increases. 


\section{Conclusions and outlook}
We have investigated the optimization of quantum networks by adjusting their topology to the desired functionality. Such an optimization is impossible in classical networks (and bottom-up quantum networks), as network connections need to be established by physical links, which in turn are limited by the underlying geometry of the network. In some cases its impossible or too costly to provide direct links between certain (long-distance) nodes. In turn, entanglement-based networks offer the possibility to establish and store entanglement shared between basically arbitrary nodes in the network, and adjust the stored states to optimize network performance w.r.t. the desired functionality. The topology of the entanglement-based network is thereby independent of the underlying physical network topology. 

We have introduced different methods and algorithms to minimize the memory requirements of the stored entangled states, thereby also reducing the amount of entanglement that is required to achieve the desired functionality. Each of the techniques we introduced is capable to significantly reduce the memory requirements, as we demonstrate with different examples. Most remarkable, even in some generic cases a scaling advantage can be obtained, reducing the required number of stored qubits per node from linearly growing with the network size to a constant. Furthermore, the techniques we developed can be combined and applied at different nested layers to further enhance the performance.

First, we adapt spectral clustering techniques from graph theory that allow us to identify homogeneously connected nodes and treat them independently. This leads to beneficial resource constructions with favourable and simpler entanglement requirements.  In addition, we introduced a merging algorithm that allows one to identify minimal multipartite resource states that allow one to fulfill all requests from some set. We also analyzed and proposed generalized constructions for efficient resource states that allow for full pairwise connectivity, i.e. arbitrary simultaneous pairwise connections. Each of the protocols and combinations of them provide powerful tools to design entanglement-based quantum networks with improved performance, and may find application beyond the specific use we considered here.

In this work we have considered a fully entanglement-based network, where the desired functionality is guaranteed by a stored entangled state, without further using the underlying quantum network. It should be emphasized, though, that also hybrid approaches are possible, where pre-established entanglement is used to improve the performance of the remaining network to establish desired target state on demand. Which types of states can optimally be used to enhance the performance of the remaining network is an interesting and relevant open question. How noise and imperfections in the generation, storage and manipulation of resource states influences the choice of state, and the performance of the schemes will be treated in future work.

\section*{Acknowledgments} This work was supported by the Austrian Science Fund (FWF) through
projects No. P30937-N27, No. P36009-N and No. P36010-N. Finanziert von der Europäischen Union - NextGenerationEU.

\bibliographystyle{apsrev4-2}
\bibliography{OptNetBiblio2}

\clearpage

\onecolumn\newpage
\appendix

\section{Problem setting} \label{Sec:Appendixswitch}
In this Appendix we provide a detailed and illustrative analysis of one example of one of the relevant approaches introduced in the main text, that can help the reader for a better understanding of the problem setting and solutions proposed in this work.

In this work we consider a top-down approach for entangled-based quantum networks. A network resource state is provided beforehand, which should be able to guarantee certain requests by local operations. A request consists of a  particular configuration of entangled states shared between the network nodes that are used to perform certain tasks (e.g. teleportation). Given $m$ different requests, the objective is to find the minimum network resource state (in terms of stored qubits) that can be transformed to any of the $m$ requests with local operations. 

Throughout the work we make two assumptions on the kind of request considered, i.e.
\begin{itemize}
    \item  In each request only bipartite links or connections, i.e. Bell pairs, can appear.
    \item In each request, each node can be at most connected to another one.
\end{itemize}
We remark that these assumptions are considered for the sake of simplicity, but all the protocols and strategies introduced can be applied for arbitrary requests with arbitrary multipartite entangled connections. 

As an example, consider a network with $n=4$ nodes. Under the previous restrictions, there exist a total of $m=9$ possible requests (see Fig. \ref{fig:appendixa}.a). In some settings, only a few of them can be required, while others do not have to be guaranteed. Note that if one can obtain any of the request with two parallel links from the resource state (the 3 bottom figures in Fig. (\ref{fig:appendixa}.a)) , one can obtain any of the $m=9$ requests. In this case (full pairwise connectivity of Sec. \ref{sec:Full connectivity protocols}), the problem reduces to finding the minimum state that can guarantee the $3$ requests with two simultaneous connections. We show next how the switch-type construction can guarantee any of the request.

\section{Switch-type construction} \label{Sec:Appendix2}
In this section we analyze in detail the case where all the requests have to be guaranteed by the resource state. In particular, we analyze the so-called switch-type solution (Sec. \ref{sec:Full connectivity protocols}).

\begin{figure}[h]
\includegraphics[width=\textwidth]{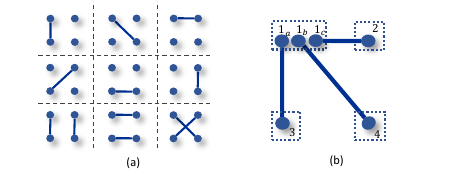}
\caption[h!]{\label{fig:appendixa} (a) All the possible target requests for a $4-$node network under our assumptions. (b) Switch-type construction that can guarantee any of the request with local operations.}
\end{figure}

The switch-type approach, see Fig. (\ref{fig:appendixa}.b), consists of three Bell states with node $1$ storing half of the qubits of each Bell state. Fig. \ref{fig:appendixa2} shows the steps required to transform this resource state into each of the $3$ aforementioned requests  with two parallel links. Note that multi-qubit operations are required, although they are local operations within node $1$. In particular, Bell state measurements ($M_{\text{Bell}}$), merging measurements ($M_{\text{Mer}}$), local complementation ($LC$) and Z measurements are used (see Sec. \ref{sec:basicops}).

An important difference with respect to other approaches introduced in this work, such as the GHZ-type construction or the merging algorithm, is the fact that the switch-type approach requires multi-qubit operations at some nodes, while in the other approaches only single qubit operations are used to transform states and fulfill the requests. Moreover, the dependence of the network on one node is another disadvantage of the switch approach. When the assumption on the requests are relaxed, and multipartite or multi-connection links are allowed in the requests, the switch-type approach is not longer favorable, while e.g. the merging algorithm is directly applicable. On the other hand, the switch approach is the optimal construction for full pairwise connectivity of a $4-$node network under the given assumptions. This can be directly seen from the entanglement shared with respect to each bipartition. 

\subsection{Optimality for four nodes}\label{Sec:Appendix3}
\begin{figure}
\includegraphics[width=\textwidth]{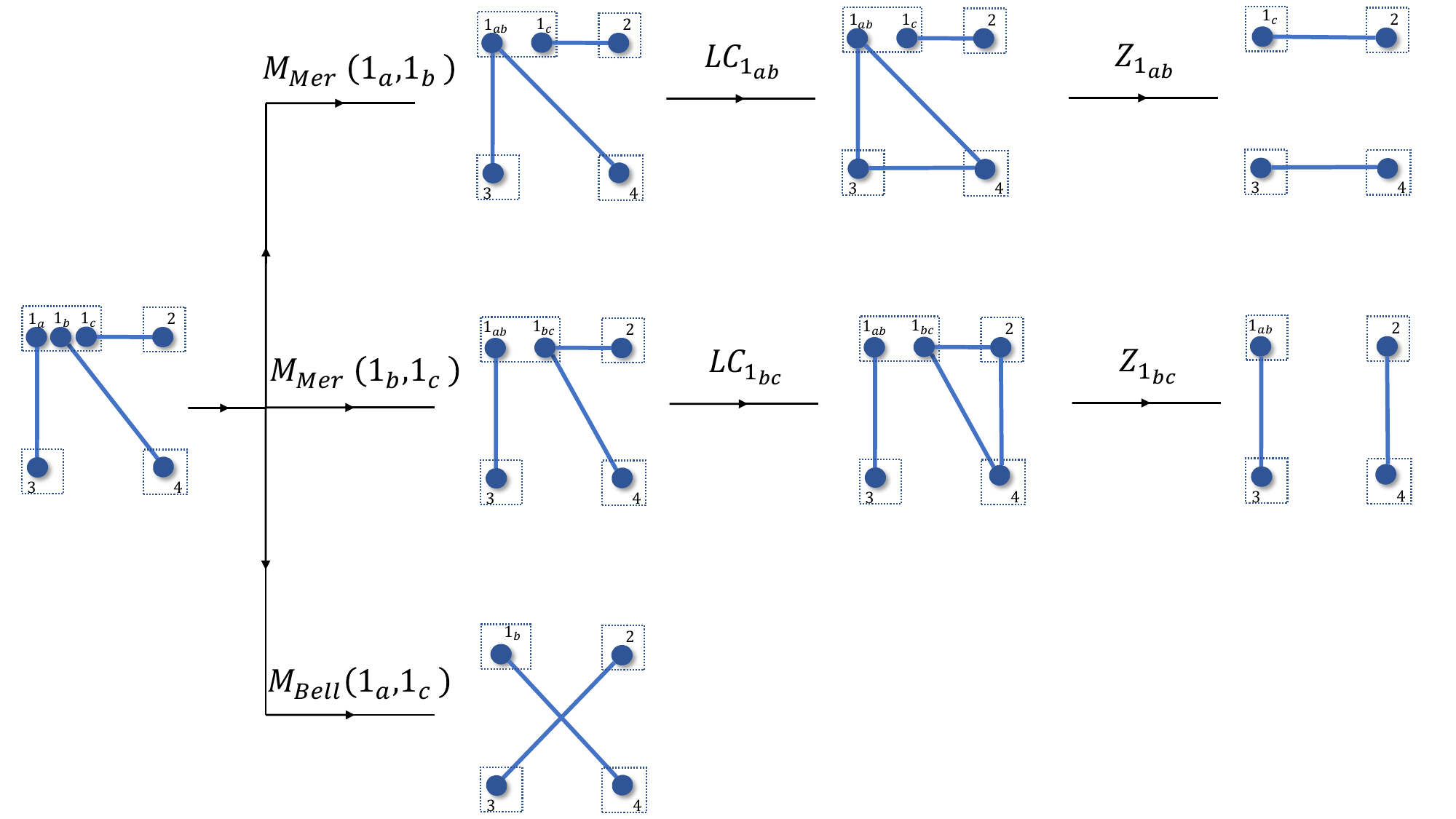}
\caption[h!]{\label{fig:appendixa2} Operations required to achieve any simultaneous requests from the switch-type construction for a network with $4$ nodes.}
\end{figure}

The Schmidt measure \cite{Schmidt2001} is a measure of entanglement that characterizes the amount of entanglement present in some multipartite graph state. The unit of bipartite entanglement given by this measure, i.e. the amount of entanglement contained in a Bell state, is called ebit. In \cite{Briegel20044} it is shown that the Schmidt measure is an entanglement monotone, and therefore cannot be increased by local operations and classical communication. In particular, this also applies with respect to any bipartition of the state. If one analyze the switch-type construction of Fig. (\ref{fig:appendixa}.b), one can see that there exist $2$ ebits of entanglement in each of the $3$ possible bipartitions of the graph. Since the simultaneous requests also involve $2$ ebits of entanglement w.r.t. each bipartition when considered together, we can conclude that no extra entanglement is spent in the process of Fig. \ref{fig:appendixa2}.  Finally, from Proposition 1 in \cite{Hahn2019}, we can conclude that there is no $5-$qubit --multipartite-- state shared by $4$ parties with at least $2$ ebits of entanglement w.r.t. each bipartition \cite{Hein2006,Briegel20044}, and then the optimality in terms of storage of the switch-type approach for $n=4$ is guaranteed.

\section{Unidirectional requests. Butterfly-type network}
\label{Sec:Appendix4}
\begin{figure}
\includegraphics[width=\textwidth]{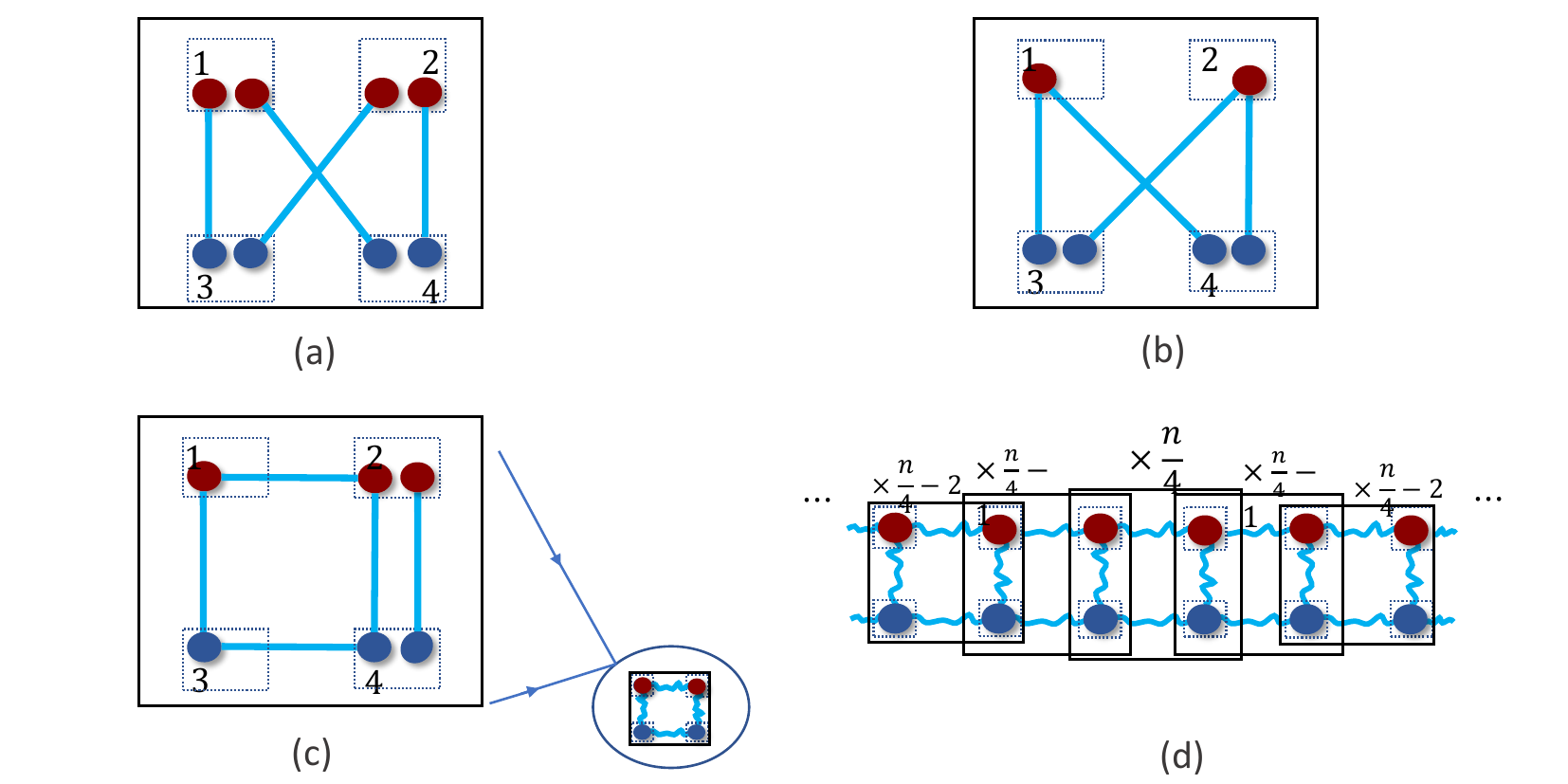}
\caption[h!]{\label{fig:appendixb1} Construction for an unidirectional network, where the request connections only involve links between red (upper) and blue (lower) nodes. (a) Direct Bell-type construction. (b) GHZ-type approach. (c) Butterfly-type construction. (d) Generalization of the butterfly-type construction for $n/2$ sender and $n/2$ receiver nodes.}
\end{figure}

A relevant network scenario consists of two sets of nodes where only connections between, but no within the sets, are required. In the example of the $4-$node network this can be seen considering the $6$ requests in Fig. (\ref{fig:appendixb1}.a) that involve connection between the two upper and two lower qubits. While clustering and merging algorithms can be directly applied for this situation, full pairwise connectivity approaches have to be adapted. Fig. \ref{fig:appendixb1} shows the full pairwise connectivity strategies introduced in the main text, adapted to this new scheme. While the basic Bell-type construction requires $\frac{n^{2}}{2}$ stored qubits, the GHZ-type approach scales as $\frac{n}{2} (\frac{n}{2}+1)$. In addition, a new possible construction exists, i.e. the butterfly-type construction.

\begin{figure}
\includegraphics[width=\textwidth]{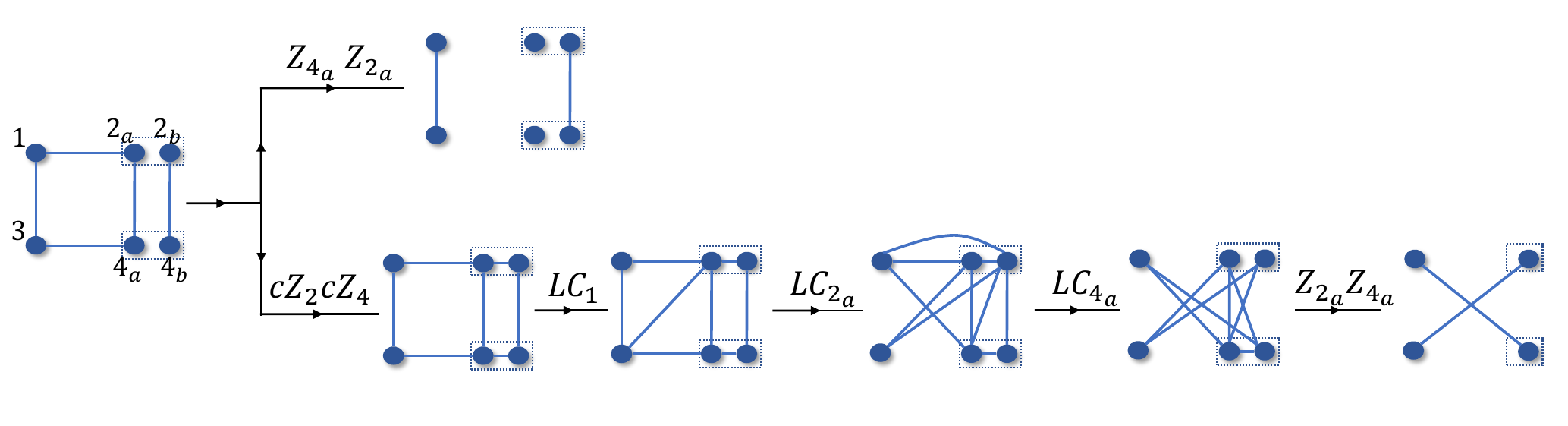}
\caption[h!]{\label{fig:appendixb2} Local operations required to reach the $2$ possible simultaneous connections between upper and lower qubits from a butterfly-type construction, in a $4-$nodes network.}
\end{figure}

The butterfly-type strategy, inspired in the results for the butterfly problem in \cite{Hahn2019}, consists of four nodes sharing a 2D cluster state, and additionally, two of them sharing an extra Bell pair (Fig. \ref{fig:appendixb1}.c). This guarantees full pairwise connectivity with only $6$ qubits in total, in opposition to the GHZ-type and to the Bell-type approach, with $7$ and $12$ total qubits respectively. Fig. \ref{fig:appendixb2} provides illustrative examples of how each request can be achieved with this network state. 

This construction can be extended and generalized for bigger networks. We propose in Fig. (\ref{fig:appendixb1}.d) a possible generalization for $n-$node networks, which requires $\frac{3}{8}n^2$ qubits in total.
This setting is particularly relevant in 2D structures with an unidirectional flow of information, where one set of nodes act as senders and the other as receivers of  the information. The extra Bell pairs can be also seen as independent nodes of the network \cite{Hahn2019}. In this sense, the presented construction can be viewed as a quantum switch between a set of input nodes and a set of output nodes.

\section{Details of the numerical analyses} \label{Sec:Numerics}
We present here further explanation for the numerical analysis of the main text, which evaluate the performance of the merging and clustering algorithms. During the numerical evaluations we restrict to the same assumptions made throughout the text, i.e. each request can only contain bipartite connections and each node can be at most connected to another one per request. 

On the one hand we analyze the performance of the merging protocol (Fig. (\ref{fig:onlymerging}.a) and Fig. (\ref{fig:onlymerging}.b)). We consider a network of $n$ nodes and we generate $m$ different random requests that we provide as input for the algorithm. The generation of the requests and the numerical evaluation can be summarized as follows. 
\begin{enumerate}
    \item A pair of nodes of the network of $n$ nodes are randomly chosen. 
    \item A link between the nodes is probabilistically generated. The probability is given by $\frac{1}{n}$.
    \item In case the link is generated, the two nodes are not longer selectable within the request. If not, the first node is kept and a new second node is randomly chosen to repeat the previous step. 
    \item When the first node link has been establish or all the potential links from that node have been rejected, repeat again with a new pair of nodes until all the nodes are taken into account.
    \item Repeat steps 1-4 for each of the $m$ requests.
    \item Provide the requests as input for the merging algorithm and compute the number of total qubits of the output resource state of the algorithm. 
    \item Repeat several rounds and compute the average number of qubits required, in order to avoid possible fluctuations arising from very favourable requests. 
\end{enumerate}

\begin{figure}
\includegraphics[width=\textwidth]{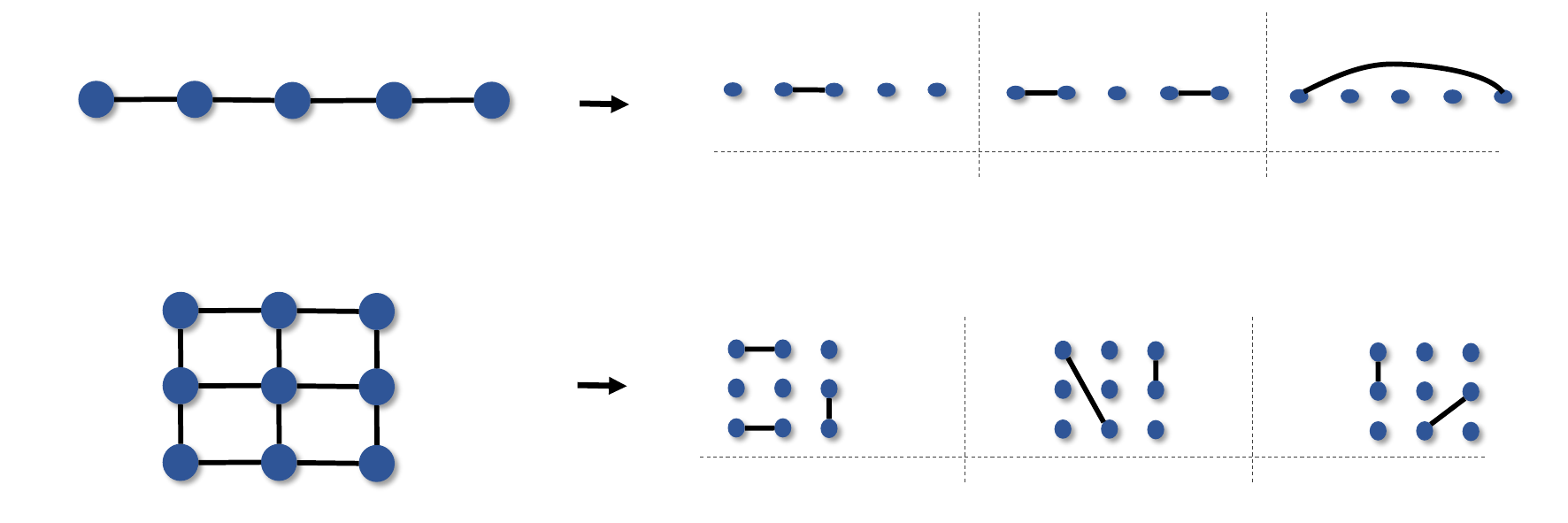}
\caption[h!]{\label{fig:appendixc} Some of the configurations that can be reached from the 1D (upper) and 2D (lower) cluster states. All of the possibilities based on the repeater-path protocol are computed and provided as input for the merging algorithm to analyze its performance.}
\end{figure}

Additionally, in order to evaluate how close to the optimal performance the merging protocol is, we consider the following scenario (Fig. (\ref{fig:onlymerging}.c) and Fig. (\ref{fig:onlymerging}.d)). Take the 1D and 2D cluster states of $n$ nodes and compute all the possible states that can be reached from them with single-qubit operations, and based on the repeater-path protocol. This consists in establishing a path between two nodes and measure the neighbor of the path in the Z basis, followed by X measurements of the path nodes. Depending on the size and location of the connections, multiple simultaneous links can appear in each request (see Fig. \ref{fig:appendixc}). In particular, for the 1D cluster state all the requests that can be obtained with local operations are considered with this method.  All of these configurations are provided as inputs for the merging algorithm (see Fig. \ref{fig:appendixc}) such that, ideally, the protocol should output exactly the cluster states, since they represent the minimum construction that can guarantee any of the requests. In particular, if we restrict to near-neighbor connections, the merging algorithm output exactly the cluster states. For large-size networks of size $n$, since the number of requests becomes intractable, we randomly choose $2n$ of the requests and repeat several runs to obtain a good estimation of how the protocol behaves. Our results (Fig. \ref{fig:onlymerging}) indicate that the merging algorithm performance is close to the optimal one. In particular, this is also the case for larger networks with increasing number of qubits. Given the NP-Complete nature of the problem, the merging protocol entails a general solution with significant efficiency.

In Fig. \ref{fig:Clusteringfigures} we analyze the clustering protocol performance and its combination with the merging algorithm. As before, given a network of $n$ nodes, we randomly generate $m$ possible target requests that the network state should be able to fulfill with local operations. For the random generation of the requests we follow the similar steps as before, with some differences, i.e. 
\begin{enumerate}
    \item A pair of nodes of the network of $n$ nodes are randomly chosen. 
    \item In this case, a link between the randomly chosen nodes is probabilistically generated but the probability depends on the network distribution. Nodes are classified in groups of certain size, and larger probability (10:1) is assigned for pair of nodes belonging to the same group. In case the nodes do not belong to the same group, the link is generated with probability $\frac{1}{n}$. Before, for the analysis of the merging protocol itself, no groups were designed and every link had the same probability ($\frac{1}{n}$) of appearing.
    \item If the link is generated, the two nodes are not longer selectable within the request. If not, the first node is kept and a new second node is randomly chosen to repeat the previous step. 
    \item When the first node link has been establish or all the potential links from that node are rejected, repeat again with a new pair of nodes until all the nodes are taken into account. Note that if all the potential links are rejected, the first node is left unconnected within that request.
    \item Repeat from each of the $m$ requests.
    \item Provide the requests as input for the clustering algorithm, which is applied once (Fig. \ref{fig:Clusteringfigures}.a) or twice (Fig. \ref{fig:Clusteringfigures}.b). Once the clustering protocol outputs the different clusters grouping the nodes according their connectivity, the merging algorithm is applied in each cluster of each layer, and we compute the total number of qubits of the overall output resource state. 
    \item Repeat several rounds and compute the average number of qubits required.
\end{enumerate}

We study the merging and clustering algorithms themselves, as well as the combined protocol of both, compared to the direct Bell-type solution to the problem (see Sec. \ref{sec:Full connectivity protocols}). A quasi-linear scaling in the storage is found with the size of the network ($n$) and the number of requests (note we study situations where the number of requests depends on the network size). This implies that each node of the network would only need to store a quasi-constant number of qubits. This effect is enhanced when the clustering algorithm is applied more times and when it is combined with the merging protocol. In addition, the clustered structure of the output state allow us to apply the merging algorithm independently for each cluster, therefore generating simpler (and therefore easier to establish and maintain) network state structures, in terms of their entanglement complexity.

\end{document}